\documentclass[twocolumn,trackchanges]{aastex631}

\usepackage{graphicx}
\usepackage{subfigure}
\usepackage{paralist}
\usepackage{amssymb}
\usepackage{bm}
\usepackage{bbding}
\usepackage[fleqn]{amsmath}
\usepackage{multirow}
\hypersetup{linkcolor=red,citecolor=blue,filecolor=cyan,urlcolor=magenta}

\submitjournal{AJ}

\shorttitle{Spin-orbit problem}
\shortauthors{Lei}

\begin{document}

\title{Dynamical structures associated with high-order and secondary resonances in the spin-orbit problem}

\correspondingauthor{Hanlun Lei}
\email{leihl@nju.edu.cn}

\author{Hanlun Lei}
\affiliation{School of Astronomy and Space Science, Nanjing University, Nanjing 210023, China}
\affiliation{Key Laboratory of Modern Astronomy and Astrophysics in Ministry of Education, Nanjing University, Nanjing 210023, China}



\begin{abstract}
In our Solar system, spin-orbit resonances are common under Sun--planet, planet--satellite and binary asteroid configurations. In this work, high-order and secondary spin-orbit resonances are investigated by taking numerical and analytical approaches. Poincar\'e sections as well as two types of dynamical maps are produced, showing that there are complicated structures in the phase space. To understand numerical structures, we adopt the theory of perturbative treatments to formulate resonant Hamiltonian for describing spin-orbit resonances. Results show that there is an excellent agreement between analytical and numerical structures. It is concluded that the main V-shape structure arising in the parameter space $(\dot\theta,\alpha)$ is sculpted by the synchronous primary resonance, those minute structures inside the V-shape region are dominated by secondary resonances and those structures outside the V-shape region are governed by high-order resonances. At last, the analytical approach is applied to binary asteroid systems (65803) Didymos and (4383) Suruga to reveal their phase-space structures.
\end{abstract}

\keywords{celestial mechanics -- minor planets, asteroids: general -- planetary systems}


\section{Introduction}
\label{Sect1}

Spin-orbit resonance is a common phenomenon in the Universe and it takes place if the rotational period of one body and its orbital period around the central object are commensurable \citep{Goldreich1966Spin, peale1977rotation}. For example, the planet Mercury is located inside the 3:2 spin-orbit resonance \citep{lemaitre20063, peale1965rotation}, and our Moon is locked inside a synchronous (1:1) spin-orbit resonance, resulting in the fact that its one side is permanently facing our Earth \citep{peale1969generalized}. Additionally, chaotic rotations are observed for minor objects in our Solar system, such as Saturn's satellite Hyperion \citep{wisdom1984chaotic}, Nyx and Hydra \citep{correia2015spin, showalter2015resonant}. In recent years, it is found that spin-orbit resonance commonly exists in binary asteroid systems \citep{pravec2016binary,naidu2015near, scheeres2006dynamical,pravec2019asteroid}.

The conventional model of spin-orbit resonance is formulated under the assumption that the two bodies move around their barycenter in fixed Keplerian orbits \citep{Goldreich1966Spin,Celletti1990AnalysisI,Celletti1990AnalysisII}. Under the classical model, spin-orbit resonance has been extensively explored. For Saturn's satellite Hyperion, \citet{wisdom1984chaotic} studied the mechanism for the onset of chaos through the overlap of first-order resonances \citep{chirikov1979universal} and they predicted that there is a large-scale chaotic behavior if the asphericity parameter $\alpha$ satisfies
\begin{equation*}
\alpha > \alpha_c = \frac{1}{2+\sqrt{14 e}}.
\end{equation*}
Under a nearly-integrable Hamiltonian model, \citet{celletti2000hamiltonian} numerically explored the stability of periodic orbits related to spin-orbit resonances by checking the eigenvalues of the monodromy matrix. \citet{flynn2005second} formulated second-order Hamiltonian model for describing spin-orbit primary resonances (not secondary resonances) by taking Lie-series transformation theory. \citet{jafari2015widespread, jafari2016chirikov} provided explicit expressions for primary spin-orbit resonance width in terms of asphericity and eccentricity and updated the Chirikov criterion for the onset of chaos. In addition, they thoroughly explored the effects of asphericity and eccentricity upon the dynamical structures by taking advantage of Poincar\'e sections. Later, \citet{nadoushan2016geography} provided a geography of varieties of rotational resonances under a fourth-order Hamiltonian model. 

If the orbit and rotation of a binary are coupled, it becomes the spin-orbit coupling problem. In this respect, \citet{naidu2015near} simulated the coupled spin and orbital motions for binary asteroid systems using Poincar\'e surfaces of section. \citet{hou2017mutual} provided explicit expressions to compute the mutual potential, force and torque between two rigid bodies with arbitrary shapes and mass distributions. Under the ellipsoid-ellipsoid model, \citet{hou2017note} presented modified Hamiltonian models for describing spin-orbit, spin-spin and spin-orbit-spin resonances, showing that the spin-orbit resonance center may change with the mass ratio and mutual distance and the center of spin-spin and spin-orbit-spin resonances may change with the rotation states of the binary asteroid. Recently, \citet{jafari2023surfing} offered a criterion for dynamical closeness of asteroid pair based on the difference between rotational and orbital angular momenta, and the author explored the dependence of phase-space structures upon the semimajor axis, eccentricity, mass ratio of the secondary to the primary and equatorial elongation of the secondary. 

Almost all the works mentioned above concentrate on primary spin-orbit resonances. However, if the frequency of small-amplitude oscillations about the primary resonance and the orbital frequency are commensurable, secondary spin-orbit resonances may happen. The existence of secondary resonances make the phase-space structures be more complicated. In this respect, \citet{wisdom2004spin} developed a perturbative model to study the secondary 3:1 spin-orbit resonance of Enceladus and discussed the rate of tidal dissipation for Enceladus inside this secondary resonance. \citet{gkolias2016theory,gkolias2019accurate} took advantage of Lie-series transformation theory to formulate analytical approximations for those low-order secondary resonances (1:1, 2:1 and 3:1) bifurcating from the synchronous primary resonance. For the bifurcation diagram of secondary 1:1 resonance, the analytical (normal form) solution can agree with numerical result in the range of $\alpha\in [0, 1.2]$, but starts to diverge when $\alpha$ is greater than 1.2 \citep{gkolias2019accurate}. 

It is not difficult to see that the Hamiltonian model based on Lie-series transformation adopted by \citet{flynn2005second} can describe high-order primary resonances outside the primary island but cannot describe secondary resonances inside the primary island, while the Hamiltonian model based on Lie-series method adopted by \citet{gkolias2016theory,gkolias2019accurate} can be used to study secondary resonances but cannot be used to study those high-order spin-orbit resonances. Thus, it is required a unified resonant Hamiltonian model for describing high-order and secondary resonances in the spin-orbit problem to explain phase-space structures.

In this work, two types of dynamical maps (the fast Lyapunov indicator and the normalized second-derivative-based index) are produced in the $(\dot\theta,\alpha)$ space. There are distinguishable structures, governed by synchronous primary resonance, high-order resonances outside the primary resonance and secondary resonances inside the primary resonance. To understand these numerical structures, the theory of perturbative treatments \citep{henrard1986perturbation,henrard1990semi} is taken to formulate a unified resonant Hamiltonian model, which can be used to study both the high-order and secondary resonances analytically. Results show that there is an excellent correspondence between analytical structures arising in phase portraits and numerical structures arising in dynamical maps.

The remaining part of this work is organized as follows. In Section \ref{Sect2}, the Hamiltonian function of the spin-orbit problem is briefly introduced. In Section \ref{Sect3}, two types of dynamical maps are produced and Section \ref{Sect4} takes advantage of the theory of perturbative treatments to formulate the resonant Hamiltonian model. Applications to binary asteroid systems are presented in Section \ref{Sect5} and conclusions are summarized in Section \ref{Sect6}.

\section{Hamiltonian model}
\label{Sect2}

The spin-orbit problem states that a triaxial satellite moves around the central planet on a fixed Keplerian orbit with semimajor axis $a$ and eccentricity $e$ \citep{murray1999solar}. To simplify the problem, we consider a planar configuration where the satellite's spin-axis (along its shortest physical axis) is assumed to be normal to its orbital plane. The principal moments of inertia of the triaxial satellite are denoted by $A$, $B$ and $C$, satisfying $A\le B\le C$. For convenience, we take the orbital semimajor axis $a$ and the mass of the central planet as the units of length and mass, respectively, and take the time unit to make the mean motion $n$ be unitary. In normalized units, the orbital period of satellite is equal to $2\pi$ and the gravitational constant ${\cal G}$ becomes unitary. 

The normalized Hamiltonian, governing the evolution of satellite's spin axis, can be written as \citep{Goldreich1966Spin, flynn2005second, gkolias2016theory, gkolias2019accurate, Celletti1990AnalysisI, Celletti1990AnalysisII}
\begin{equation}\label{Eq1}
{\cal H} = \frac{1}{2}{\dot \theta ^2} - \frac{{{\alpha ^2}}}{{4{r^3}}}\cos \left[ {2\left( {\theta  - f} \right)} \right],
\end{equation}
where $\theta$ is the angle between the satellite's longest axis and a reference line, $f$ is the true anomaly and $r$ is the radial distance from the planet and the asphericity parameter $\alpha$ is characterized in terms of the moments of inertia (or equatorial elongation of satellite) as follows \citep{murray1999solar}:
\begin{equation*}
\alpha  = \sqrt {\frac{{3\left( {B - A} \right)}}{C}} = \sqrt {\frac{{3\left( {a_s^2/b_s^2 - 1} \right)}}{{a_s^2/b_s^2 + 1}}},
\end{equation*}
where $a_s$ and $b_s$ are the semi-axis of triaxial satellite ($\alpha$ is independent upon the shortest semi-major axis $c_s$). As $r(t)$ and $f(t)$ are functions of time $t$, the Hamiltonian (\ref{Eq1}) determines a non-autonomous dynamical model. To make it be autonomous, it is standard to augment the phase space by introducing an action variable $T$ which is conjugated to time $t$. The resulting new set of canonical variables are denoted by \citep{flynn2005second}
\begin{equation*}
\begin{aligned}
q_1 = \theta,\quad p_1 = \dot \theta,\\
q_2 = t,\quad p_2 = T
\end{aligned}
\end{equation*}
and the augmented Hamiltonian becomes
\begin{equation}\label{Eq2}
{\cal H} = \frac{1}{2}{\dot q_1 ^2} + p_2 - \frac{{{\alpha ^2}}}{{4{r^3}}}\cos \left[ {2\left( {q_1  - f} \right)} \right],
\end{equation}
which determines a 2-DOF dynamical model. In practice, it is usual to take a certain value of $p_2$ to make the Hamiltonian ${\cal H}(q_1,p_1,q_2,p_2)$ be equal to zero. 

According to elliptic expansions, the Hamiltonian (\ref{Eq2}) can be expanded in Fourier series \citep{Celletti1990AnalysisI, Celletti1990AnalysisII}. As a consequence, the spin-orbit Hamiltonian truncated at the fourth order in $e$ takes the form:
\begin{equation}\label{Eq3}
{\cal H} = \frac{{p_1^2}}{2} + {p_2} - \frac{{{\alpha ^2}}}{4}\sum\limits_{n =  - 6}^2 {{{\cal C}_n (e)}\cos \left( {2{q_1} + n{q_2}} \right)},
\end{equation}
where the coefficients are related to eccentricity in the following form:
\begin{equation*}
\begin{aligned}
{{\cal C}_{-6}} &= \frac{{533}}{{16}}{e^4},\;{{\cal C}_{-5}} = \frac{{845}}{{48}}{e^3},\;{{\cal C}_{-4}} = \frac{{17}}{2}{e^2} - \frac{{115}}{6}{e^4},\\
{{\cal C}_{-3}} &= \frac{7}{2}e - \frac{{123}}{{16}}{e^3},\;{{\cal C}_{-2}} = 1 - \frac{5}{2}{e^2} + \frac{{13}}{{16}}{e^4},\\
{{\cal C}_{-1}} &=  - \frac{1}{2}e + \frac{1}{{16}}{e^3},\;{{\cal C}_0} = 0,\;{{\cal C}_1} = \frac{1}{{48}}{e^3},\;{{\cal C}_2} = \frac{1}{{24}}{e^4}.
\end{aligned}
\end{equation*}
The equations of motion can be derived from Hamiltonian canonical relations \citep{morbidelli2002modern},
\begin{equation}\label{Eq4}
\begin{aligned}
{{\dot q}_1} &= \frac{{\partial {\cal H}}}{{\partial {p_1}}},\quad {{\dot p}_1} =  - \frac{{\partial {\cal H}}}{{\partial {q_1}}},\\
{{\dot q}_2} &= \frac{{\partial {\cal H}}}{{\partial {p_2}}},\quad {{\dot p}_2} =  - \frac{{\partial {\cal H}}}{{\partial {q_2}}}.
\end{aligned}
\end{equation}
In summary, the spin-orbit problem is ruled by two parameters: the orbital eccentricity $e$ and the asphericity parameter $\alpha$. In Sect. \ref{Sect3}, the equations of motion are numerically integrated in order to produce dynamical maps as well as Poincar\'e sections.

\section{Dynamical maps}
\label{Sect3}

In this section, we numerically explore phase-space structures of the spin-orbit problem by means of dynamical maps, which provides a global view about the dynamics. In particular, two types of indicators are adopted. In producing two types of dynamical maps, the asphericity parameter $\alpha$ and the spin frequency $\dot \theta$ are distributed in the interval $[0.2,1.5]$$\times$$[-0.5,2.5]$ with steps of $\delta \alpha = 0.01$ and $\delta \dot\theta = 0.01$ and numerical integration is performed over 500 orbital periods.

The first one is the fast Lyapunov indicator (FLI), introduced by \citet{froeschle1997fast}, which provides a quick method for distinguishing between chaotic and regular regions in the phase space. There are several variants for the definition of FLI \citep{barrio2005sensitivity, barrio2006painting, skokos2009lyapunov}. In this work, we adopt the definition of FLI, given by \citep{guzzo2002numerical}
\begin{equation}\label{Eq5}
{\rm FLI} = \mathop {\sup }\limits_{t \in \left[ {{t_0},{t_f}} \right]} {\log _{10}}\frac{{\left\| {\delta {\bm X}\left( t \right)} \right\|}}{{\left\| {\delta {\bm X}\left( {{t_0}} \right)} \right\|}}
\end{equation}
where $\bm X = (\theta, \dot\theta)^{\rm T}$ is the state vector, ${\delta {\bm X}\left( {{t_0}} \right)}$ is the initial tangent vector, ${\delta {\bm X}\left( {{t}} \right)}$ is the deviation at the moment $t$, and $t_f$ is the maximum integration time. In general, the magnitude of FLI increases linearly with time for regular motions and it grows exponentially with time for chaotic motions \citep{guzzo2002numerical,skokos2009lyapunov}. As a result, the maps of FLI can clearly show regular and chaotic regions.

\begin{figure*}
\centering
\includegraphics[width=\columnwidth]{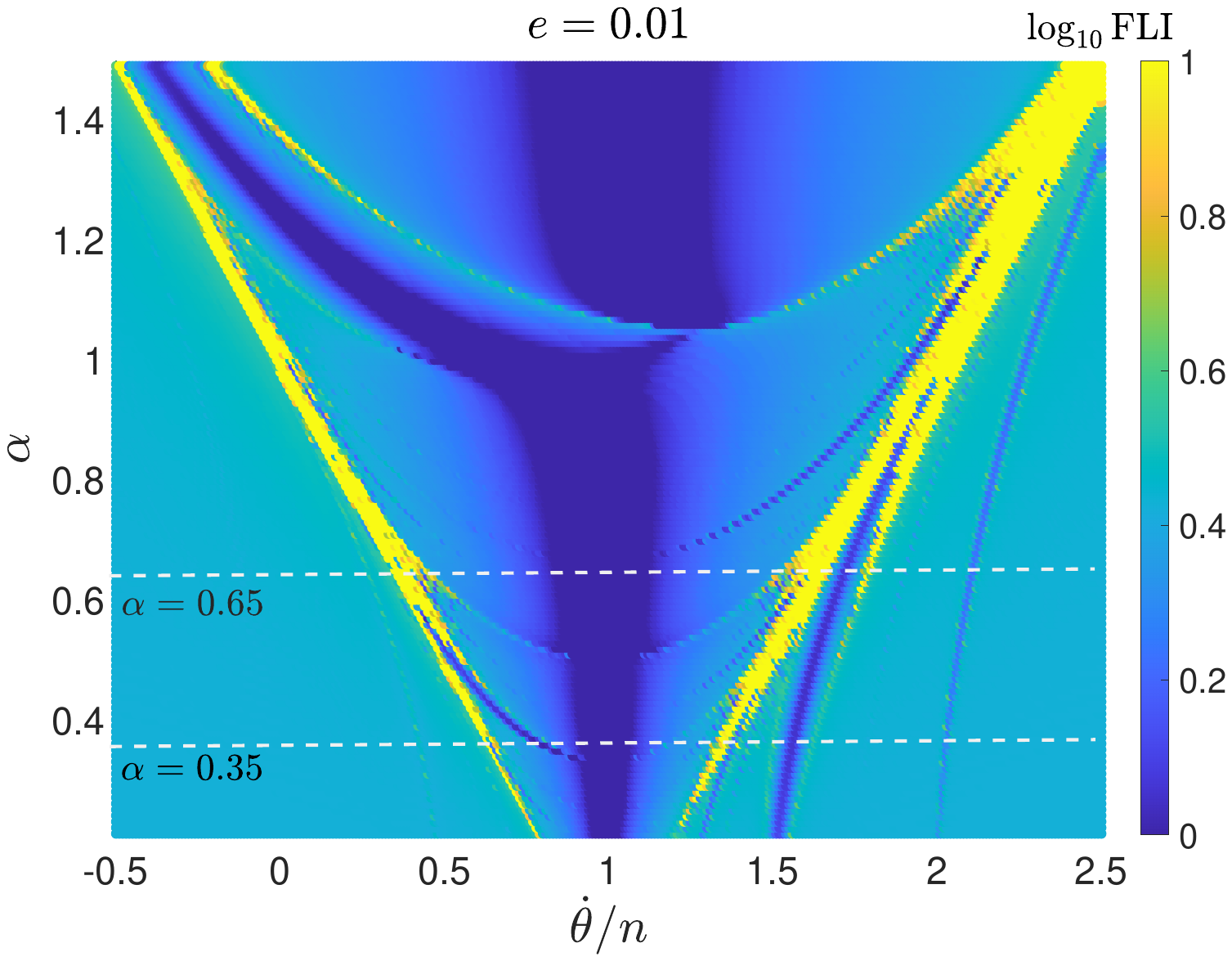}
\includegraphics[width=\columnwidth]{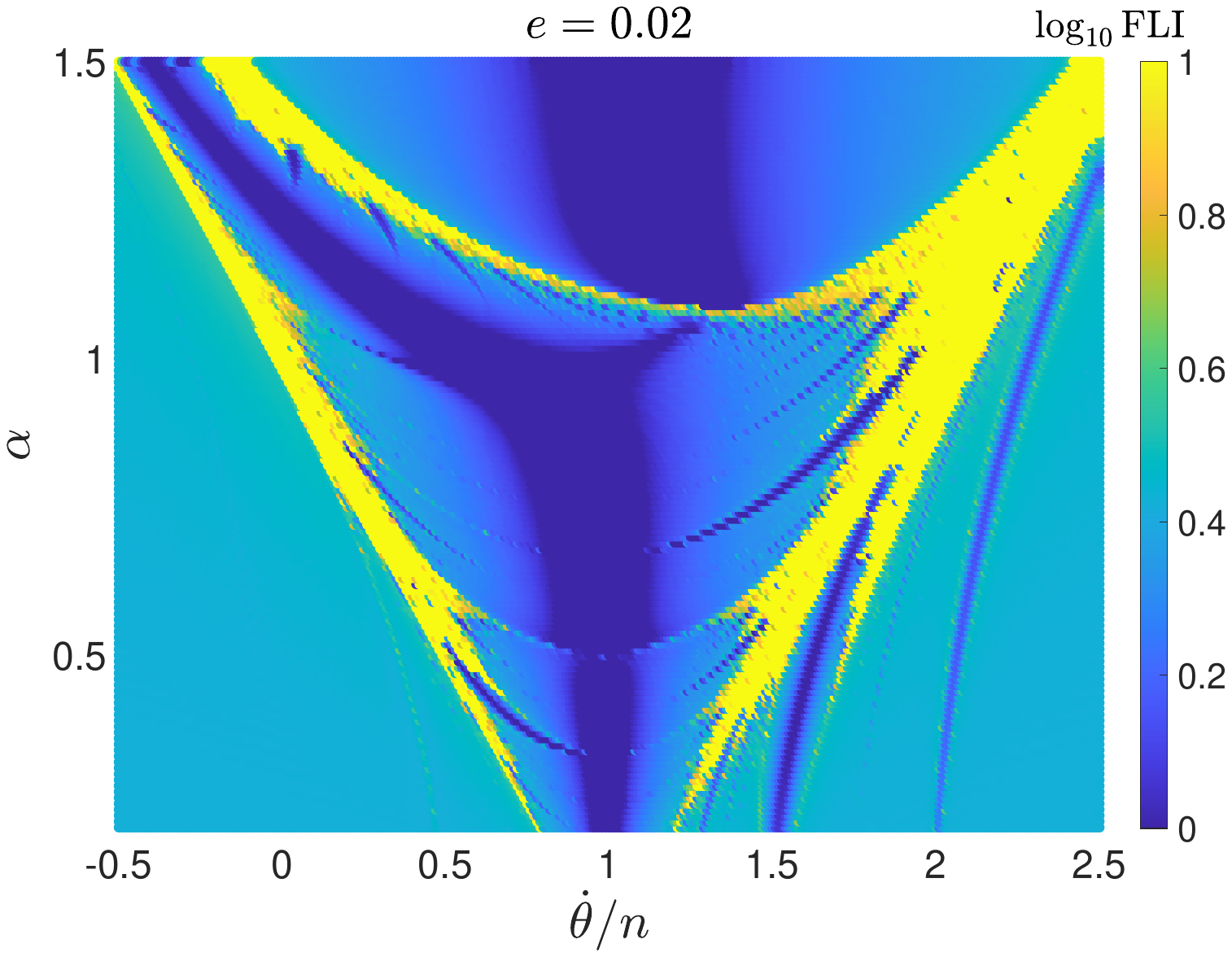}
\caption{Dynamical maps of the fast Lyapunov indicator (FLI) shown in the $(\dot \theta, \alpha)$ space for the configurations with orbital eccentricities at $e=0.01$ (\textit{left panel}) and $e=0.02$ (\textit{right panel}). The index shown in the color bar corresponds to the base 10 logarithm of FLI. In both panels, the main V-shape structures are visible. The red dashed lines stand for the locations of $\alpha = 0.35$ and $\alpha = 0.65$ and their associated Poincar\'e sections are shown in Fig. \ref{Fig3}.}
\label{Fig1}
\end{figure*}

For the spin-orbit problem, we take the maximum integration time as 100 times the orbital period and fix the initial angle at $\theta = 0$. In practical computation, we take the initial deviation relative to the reference orbit as ${\delta {\bm X}\left( {{t_0}} \right)} = \sqrt{2}/2 \times 10^{-4}$ which include a small component along the tangent vector (our experient shows that the choice of ${\delta {\bm X}\left( {{t_0}} \right)}$ has no influence upon the maps of FLI). Fig. \ref{Fig1} shows the maps of FLI in the $(\dot \theta, \alpha)$ space for two configurations with orbital eccentricities at $e=0.01$ and $e=0.02$. The blue color, standing for low magnitude of FLI, indicates regular dynamics and the yellow color, corresponding to high magnitude of FLI, indicates chaotic dynamics. From Fig. \ref{Fig1}, we can see that (a) the structures corresponding to $e=0.01$ and $e=0.02$ are qualitatively similar and the latter one has wider chaotic layers than the former one, (b) the chaotic strips in both panels presents V-shape structure in the $(\dot \theta, \alpha)$ space, (c) inside the V-shape region some sub-structures caused by secondary resonances can be visibly observed and (d) outside the V-shape region additional sub-structures caused by high-order spin-orbit resonances are visible. As shown in the following sections, the V-shape strips arising in the maps of FLI are due to the dynamical separatrix of the synchronous primary resonance. Around the separatrix, chaotic layers form the main V-shape structures. 

The second indicator we adopt is the second-derivative-based index $\left\| {\Delta D} \right\|$ where $D$ measures the diameter of an orbit, developed by \citet{Daquin2023Detection}. The second-derivative-based index is a sensitive and robust indicator, which can be used to detect separatrices, resonant webs and chaotic seas in the phase space. In particular, such an indicator has a strong ability to distinguish minute structures in the phase space. 

For the spin-orbit problem, the action-angles variables are denoted by $(\theta,\dot\theta)$ and the diameter metric of an orbit is defined by
\begin{equation}\label{Eq6}
D\left( {{\theta_0},{\dot\theta _0}} \right) = \mathop {\max }\limits_{{t_0} \le t \le {t_f}} {\dot\theta}\left( {{\theta_0},{\dot\theta _0},t} \right) - \mathop {\min }\limits_{{t_0} \le t \le {t_f}} {\dot\theta}\left( {{\theta_0},{\dot\theta _0},t} \right)
\end{equation}
where $(\theta_0,\dot\theta_0)$ stand for the initial state and $t_0$ and $t_f$ are the initial and final moments of time. Based on the distance metric $D$, a normalized second-order-derivative-based quantity is introduced by
\begin{equation}\label{Eq7}
\left\| {\Delta D} \right\| = \frac{1}{D\left( {{\theta_0},{\dot\theta _0}} \right)}\left(
\left|\frac{\partial^2 D\left( {{\theta_0},{\dot\theta _0}} \right)}{\partial {\dot\theta_0^2}}\right|
+\left|\frac{\partial^2 D\left( {{\theta_0},{\dot\theta _0}} \right)}{\partial {\theta_0^2}}\right|
\right).
\end{equation}
It should be mentioned that the introduction of the index $\left\| {\Delta D} \right\|$ follows from recent developments on Lagrangian descriptors and arc-length of orbits \citep{daquin2022global}. Regarding the computation method (central difference) about the second derivatives of the diameter, please refer to \citet{Daquin2023Detection} for more details. Contrarily to the FLI method, the method based on $\left\| {\Delta D} \right\|$ is free of tangent dynamics.

The $\left\| {\Delta D} \right\|$ metric allows to sharply detect the separatrix crossing and thus the maps of $\left\| {\Delta D} \right\|$ are able to provide more details on minute structures in the phase space. The difference of $\left\| {\Delta D} \right\|$ defined in this work from the conventional definition given in \citet{Daquin2023Detection} lies in the normalization. Our experiments show that the normalized indicator can detect those structures with small strength.

\begin{figure*}
\centering
\includegraphics[width=\columnwidth]{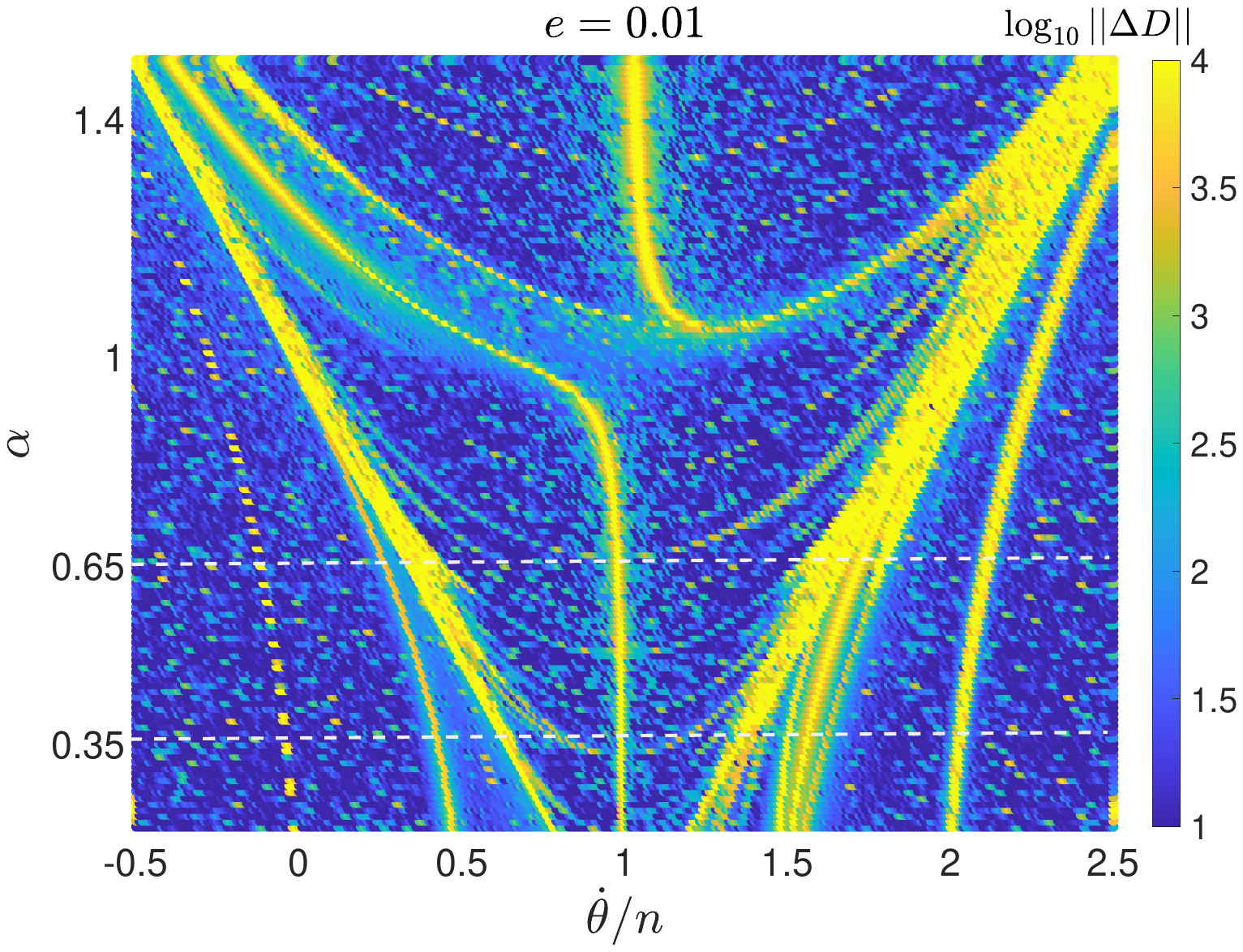}
\includegraphics[width=\columnwidth]{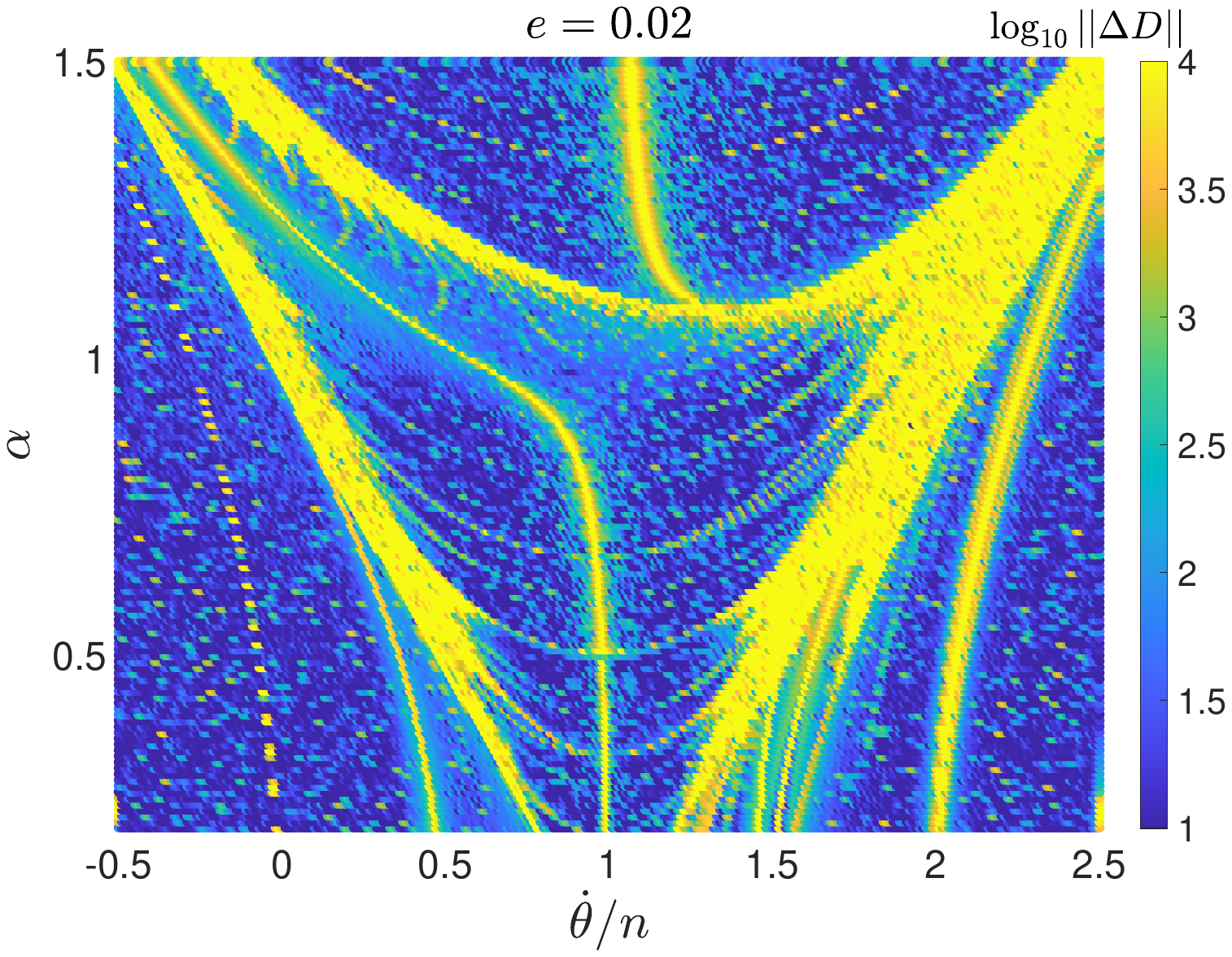}
\caption{Dynamical maps of the second-derivative-based index shown in the $(\dot \theta, \alpha)$ space for the configurations with orbital eccentricities at $e=0.01$ (\textit{left panel}) and $e=0.02$ (\textit{right panel}). The index shown in the color bar corresponds to the base 10 logarithm of the normalized second-derivative-based index $||\Delta D||$ (to make the structures be distinguishable, the index $||\Delta D||$ is limited in the interval [10,10000]). In both panels, the main V-shape structures are visible. Compared to the maps of FLI shown in Fig. \ref{Fig1}, minute structures can be visibly observed, especially in the region on the left-hand side of the V-shape structure. Poincar\'e sections corresponding to the cases of $\alpha = 0.35$ and $\alpha=0.65$ are presented in Fig. \ref{Fig3}.}
\label{Fig2}
\end{figure*}

For the spin-orbit problem, we take the maximum integration time as 100 times the orbital period and fix the initial angle at $\theta_0 = 0$ (this setting is the same as that made in FLI maps). Fig. \ref{Fig2} presents the dynamical structures of the normalized second-derivative-based index $||\Delta D||$ in the $(\dot\theta,\alpha)$ space for eccentricities at $e=0.01$ and $e=0.02$. Besides the V-shape structures, some minute structures can be visibly detected. In particular, the structures on the left side of the V-shape structure can be visibly observed in the maps of $||\Delta D||$, while they are not visible in the maps of FLI. In addition, more sub-structures can be found inside the V-shape region.

\begin{figure*}
\centering
\includegraphics[width=\columnwidth]{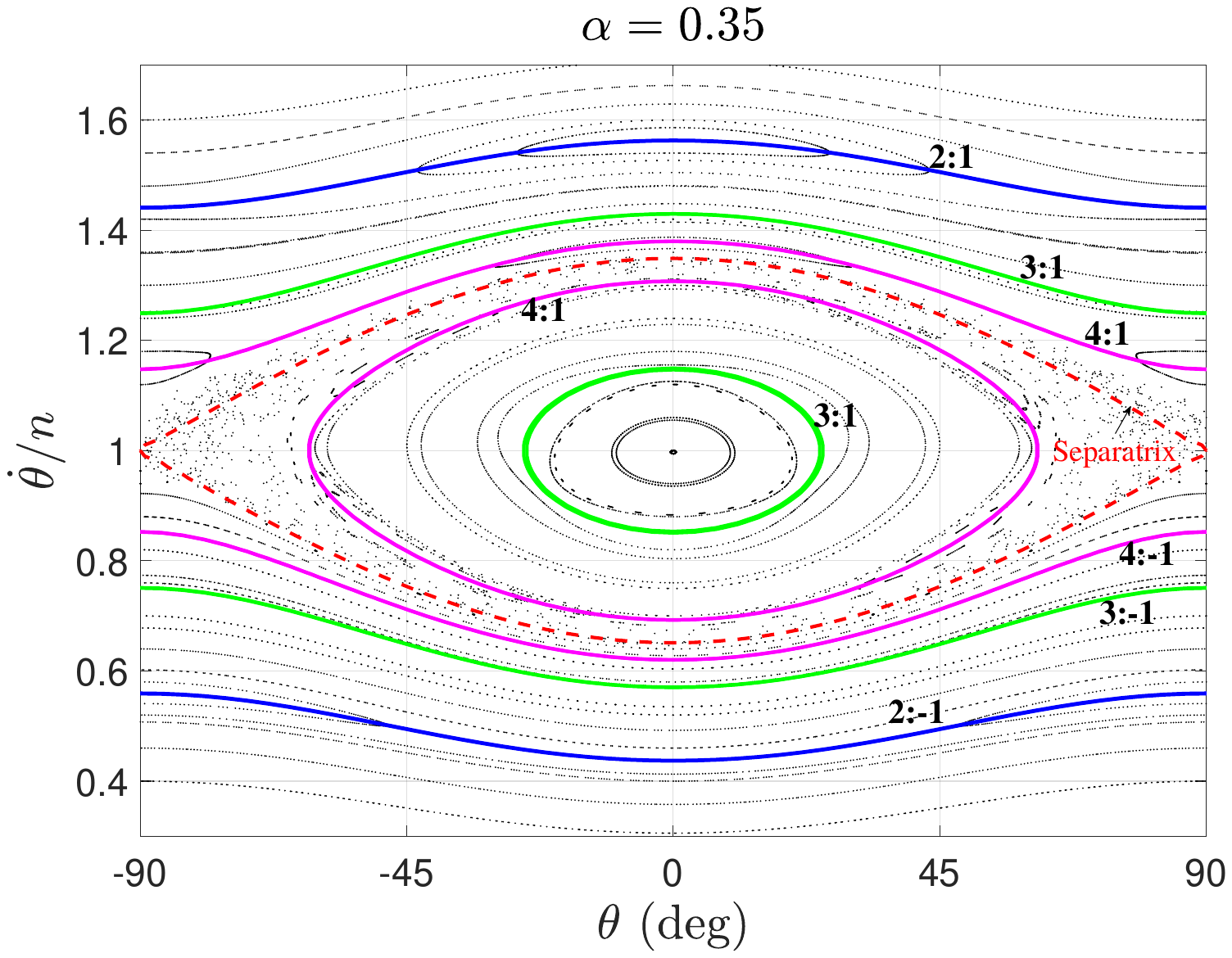}
\includegraphics[width=\columnwidth]{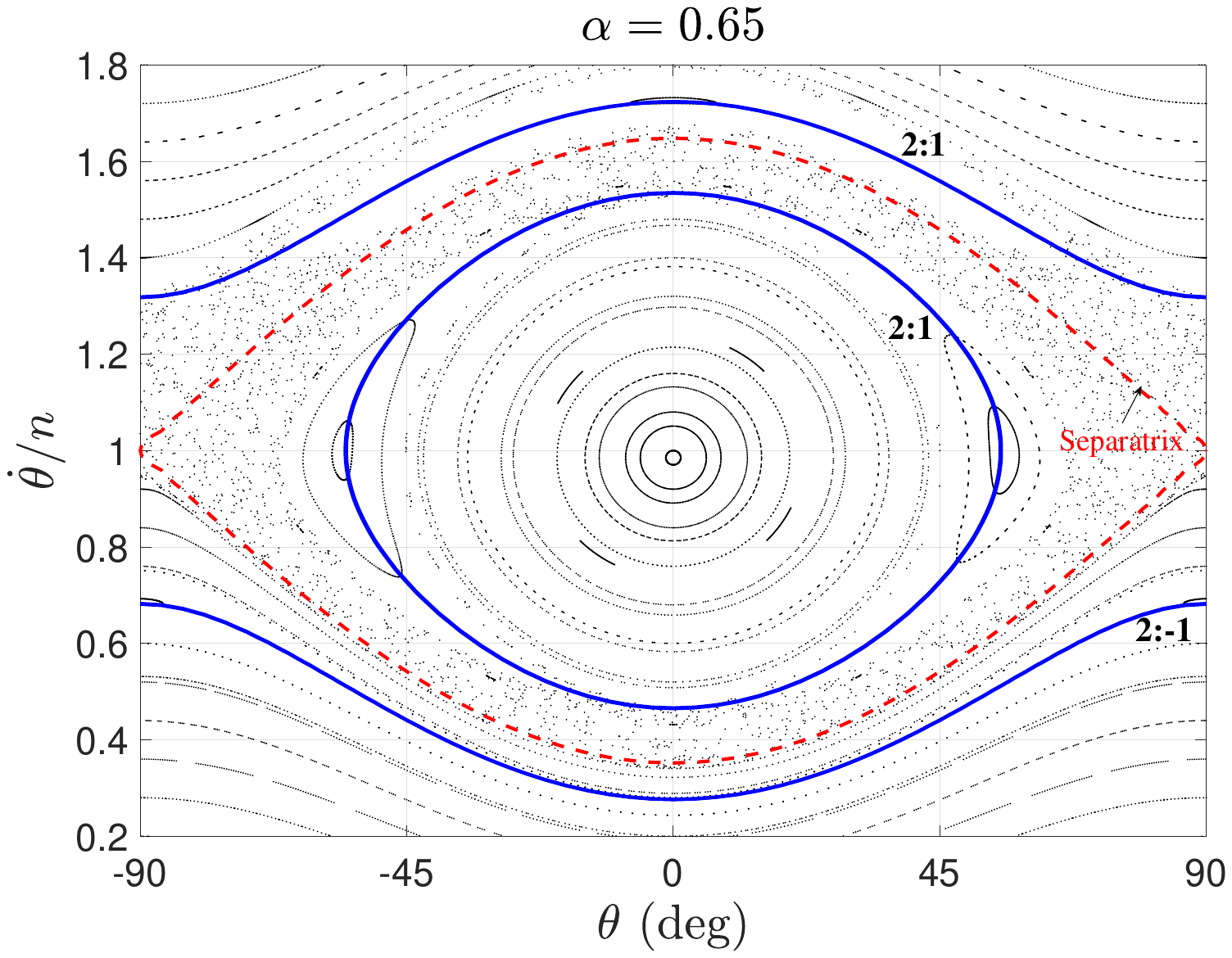}
\caption{Nominal location of the centers of high-order and secondary resonances determined under the Kernel Hamiltonian model together with the Poincar\'e surfaces of section for spin-orbit problems with asphericity parameters at $\alpha = 0.35$ (\textit{left panel}) and $\alpha = 0.65$ (\textit{right panel}). Nominal locations of high-order and secondary resonances are determined by equation (\ref{Eq14}). The eccentricity is taken as $e=0.01$. The red dashed lines stand for the dynamical separatrix of the synchronous primary resonance, specified by equation (\ref{Eq10}). The primary resonance happens inside the region bounded by the dynamical separatrices.}
\label{Fig3}
\end{figure*}

To understand the structures arising in the dynamical maps shown in Figs. \ref{Fig1} and \ref{Fig2}, we take two examples with $\alpha = 0.35$ and $\alpha=0.65$, which are explicitly marked in the left panels. To produce Poincar\'e sections, we record the states every time when the variable $q_2$ (or $t$) satisfies
\begin{equation*}
{\rm mod}\left({q_2,2\pi}\right) = {\rm mod}\left({t,2\pi}\right) = 0.
\end{equation*}
In simulation, the state variable $p_2$ is chosen to ensure that $\cal H$ is equal to zero and the numerical integration is taken over 500 orbital periods. Poincar\'e sections of these two representative systems are produced in the case of $e=0.01$, as shown in Fig. \ref{Fig3}. From the Poincar\'e sections, we can see that (a) in the case of $\alpha = 0.35$, the secondary 3:1 and 4:1 spin-orbit resonances can be observed and (b) in the case of $\alpha = 0.65$ only the secondary 2:1 resonance can be found inside the synchronous primary resonance. These secondary resonance are responsible for the structures arising inside the V-shape region shown in Figs. \ref{Fig1} and \ref{Fig2}. In both panels of Fig. \ref{Fig3}, we can further observe (a) chaotic layers around the separatrices of the primary resonance, corresponding to the V-shape main structures, and (b) high-order resonances outside the primary resonance, corresponding to the sub-structures outside the V-shape region.

From the dynamical maps shown in Figs. \ref{Fig1} and \ref{Fig2}, we may ask: how do high-order and/or secondary spin-orbit resonances sculpt the dynamical structures arising in the phase space? In the coming section, we aim at addressing this problem by taking advantage of the theory of perturbative treatments developed by \citet{henrard1986perturbation} and \citet{henrard1990semi}. 

\section{Perturbative treatments}
\label{Sect4}

To describe the synchronous primary resonance, we introduce the following canonical transformation:
\begin{equation*}
\begin{aligned}
{\sigma _1} &= {q_1} - {q_2},\quad {\Sigma _1} = {p_1},\\
{\sigma _2} &= {q_2},\quad {\Sigma _2} = {p_2} + {p_1},
\end{aligned}
\end{equation*}
where the argument $\sigma_1$ is the resonant angle of the synchronous primary resonance. Under such a new set of variables, the Hamiltonian can be re-organized as follows:
\begin{equation}\label{Eq8}
\begin{aligned}
{\cal H} = & \frac{{\Sigma _1^2}}{2} +  {{\Sigma _2} - {\Sigma _1}}  - \frac{{{\alpha ^2}}}{4}{{\cal C}_{ - 2}}\left( e \right)\cos \left( {2{\sigma _1}} \right)\\
&- \frac{{{\alpha ^2}}}{4}\sum\limits_{\scriptstyle - 6 \le n \le 2\hfill \atop
\scriptstyle n \ne  - 2\hfill} {{{\cal C}_n}\left( e \right)\cos \left[ {2{\sigma _1} + \left( {n + 2} \right){\sigma _2}} \right]}.
\end{aligned}
\end{equation} 

From the viewpoint of perturbative treatment \citep{henrard1986perturbation, henrard1990semi}, the Hamiltonian is composed of the kernel Hamiltonian (unperturbed part) and the part of perturbation, denoted by
\begin{equation}\label{Eq9}
{\cal H} = {\cal H}_0 \left(\sigma_1,\Sigma_1,\Sigma_2\right) + {\cal H}_1\left(\sigma_1,\Sigma_1,\sigma_2,\Sigma_2\right),
\end{equation}
where the unperturbed part is
\begin{equation*}
{{\cal H}_0} = \frac{{\Sigma _1^2}}{2} +  {{\Sigma _2} - {\Sigma _1}}  - \frac{{{\alpha ^2}}}{4}\cos \left( {2{\sigma _1}} \right)
\end{equation*}
and the perturbation part is
\begin{equation*}
\begin{aligned}
{\cal H}_1 = & - \frac{{{\alpha ^2}}}{4}\left[{{\cal C}_{ - 2}}\left( e \right) - 1\right]\cos \left( {2{\sigma _1}} \right)\\
&- \frac{{{\alpha ^2}}}{4}\sum\limits_{\scriptstyle - 6 \le n \le 2\hfill \atop
\scriptstyle n \ne  - 2\hfill} {{{\cal C}_n}\left( e \right)\cos \left[ {2{\sigma _1} + \left( {n + 2} \right){\sigma _2}} \right]}.
\end{aligned}
\end{equation*}
The reason why we keep those periodic terms related to $\sigma_1$ in the kernel Hamiltonian  is  that the magnitudes of coefficients related to $\sigma_1$ are at the zeroth order in $e$. In this way, the influence of synchronous primary resonance upon high-order and/or secondary resonances are taken into consideration.   

The dynamical model specified by the kernel Hamiltonian ${\cal H}_0$ is of one degree of freedom and it can be used to describe the dynamical structure of the synchronous primary resonance. Under the perturbation of ${\cal H}_1$, high-order spin-orbit resonances (outside the primary resonance) as well as secondary resonances (inside the primary resonance) may appear in the phase space.

In this section, we will first study the basic dynamics governed by the kernel Hamiltonian and determine the nominal location of high-order and secondary resonances in Sect. \ref{Sect4-1}, then formulate resonant Hamiltonian models by taking advantage of perturbative treatments \citep{henrard1986perturbation, henrard1990semi} in Sect. \ref{Sect4-2} and results are shown in Sect. \ref{Sect4-3}.

\subsection{Dynamics under the Kernel Hamiltonian model}
\label{Sect4-1}

As $\sigma_2$ is absent from ${\cal H}_0$, the unperturbed Hamiltonian model holds an integral of motion $\Sigma_2$. Thus, the kernel Hamiltonian determines an integrable dynamical model. It is not difficult to get that the stable equilibrium points are located at $(2\sigma_1 = 0, \Sigma_1 = 1)$ and the unstable equilibrium points are located at $(2\sigma_1 = \pi, \Sigma_1 = 1)$. Thus, the synchronous primary resonance is centered at $(2\sigma_1 = 0, \Sigma_1 = 1)$. The level curve of Hamiltonian passing through the saddle point in the phase space, satisfying
\begin{equation*}
\frac{1}{2} \left(\Sigma _1^2 + 1\right) - {\Sigma _1} - \frac{{{\alpha ^2}}}{4} \left(\cos 2{\sigma _1} + 1\right) =0.
\end{equation*}
As a consequence, the dynamical separatrix of the synchronous primary resonance can be explicitly expressed by 
\begin{equation}\label{Eq10}
{\Sigma _1} = 1 \pm \frac{{\sqrt 2 }}{2}\alpha \sqrt {1 + \cos 2{\sigma _1}},
\end{equation}
where the symbol `+' corresponds to the upper separatrix and `-' corresponds to the lower one. Usually, the distance between the upper and lower separatrices evaluated at the resonance centre $(2\sigma_1 = 0, \Sigma_1 = 1)$ is called the resonant width. According to equation (\ref{Eq10}), the resonant width (in normalised units) is equal to
\begin{equation*}
\Delta \Sigma_1 = \Delta \dot\theta = \alpha.
\end{equation*}
It shows that the asphericity parameter $\alpha$ stands for the resonant width (or strength) of the synchronous primary resonance.

Dynamical separatrix plays an role in dividing the entire phase space into regions of circulation and libration. In particular, the region bounded by the separatrices described by equation (\ref{Eq10}) are referred to as the librating zone and the remaining regions are called circulating zones. In addition, the level curves of Hamiltonian in the phase space (i.e., phase portraits) stand for the trajectories, which are exactly periodic.

Under the unperturbed Hamiltonian model, let us introduce action--angle variables as follows \citep{morbidelli2002modern}:
\begin{equation}\label{Eq11}
\begin{aligned}
\sigma _1^{\rm{*}} &= {\sigma _1} - \rho \left( {\sigma _1^{\rm{*}},\Sigma _1^{\rm{*}},{\Sigma _2^*}} \right),\quad \Sigma _1^{\rm{*}} = \frac{1}{{2\pi }}\oint {{\Sigma _1}{\rm d}{\sigma _1}}\\
\sigma_2^* &= \sigma_2,\quad \Sigma_2^*=\Sigma_2,
\end{aligned}
\end{equation}
where $\sigma _1^{\rm{*}}$ becomes a linear function of time, $\rho \left( {\sigma _1^{\rm{*}},\Sigma _1^{\rm{*}},{\Sigma _2^*}} \right)$ is a periodic function with period $T$ and $\Sigma _1^{\rm{*}}$ is called Arnold action, which represents the signed area bounded by the level curves in the phase space divided by $2\pi$ \citep{morbidelli2002modern}. The Arnold action is equal to zero at the centre of the synchronous resonance and it reaches the maximum along the separatrix of the primary resonance.

The transformation given by equation (\ref{Eq11}) is canonical with the following generating function,
\begin{equation*}
S\left( {{\sigma _1},\Sigma _1^*,\sigma_2,\Sigma_2^*} \right) = \sigma_2 \Sigma_2^* + \int {{\Sigma _1}\left( {{{\cal H}_0}\left( {\Sigma _1^*}, \Sigma_2^* \right),{\sigma _1}} \right){\rm d}{\sigma _1}}.
\end{equation*}
After the canonical transformation, the unperturbed Hamiltonian can be written as a normal form,
\begin{equation}\label{Eq12}
{\cal H}_0\left(\sigma_1,\Sigma_1,\Sigma_2\right) \Rightarrow  {{\cal H}_0}\left( {\Sigma _1^*,\Sigma _2^*} \right)
\end{equation}
which determines the fundamental frequencies, given by
\begin{equation}\label{Eq13}
\dot \sigma _1^{\rm{*}} = \frac{{\partial {{\cal H}_0}\left( {\Sigma _1^*,\Sigma _2^*} \right)}}{{\partial \Sigma _1^*}},\quad
\dot \sigma _2^{\rm{*}} =1.
\end{equation}
In particular, when the fundamental frequencies become commensurable, high-order or secondary spin-orbit resonances may happen. As a result, the condition of a certain $k_1$:$k_2$ spin-orbit resonance takes the form,
\begin{equation}\label{Eq14}
{k_1}\dot \sigma _1^{\rm{*}} - {k_2}\dot \sigma _2^{\rm{*}} = 0 \Rightarrow {k_1}\dot \sigma _1^{\rm{*}} - {k_2} = 0.
\end{equation}
By solving equation (\ref{Eq14}), it is possible to determine the nominal location of high-order and/or secondary resonances. 

In Fig. \ref{Fig3}, the nominal locations of high-order and secondary spin-orbit resonances are shown in the phase space $(\theta, \dot\theta)$ for the spin-orbit problems with $\alpha = 0.35$ and $\alpha = 0.65$ and they are compared with the associated Poincar\'e sections. In both panels of Fig. \ref{Fig3}, the red dashed lines stand for the dynamical separatrices of the synchronous primary resonance, which are mathematically described by equation (\ref{Eq10}). It is observed that around the dynamical separatrix there is a chaotic layer, which is wilder with a higher asphericity parameter $\alpha$. 

For convenience, we refer to the spin-orbit resonances inside the primary resonance as the secondary resonances and the ones outside the primary resonance as high-order resonances. In the case of $\alpha = 0.35$, the secondary 3:1 and 4:1 resonances can be observed and, in the case of $\alpha = 0.65$, only the secondary 2:1 resonance can be seen. Comparing the distribution of high-order and secondary resonances to the Poincar\'e sections, we can further observe that the distributions of high-order and secondary resonances are in good agreement with the structures arising in Poincar\'e sections.

\begin{figure*}
\centering
\includegraphics[width=\columnwidth]{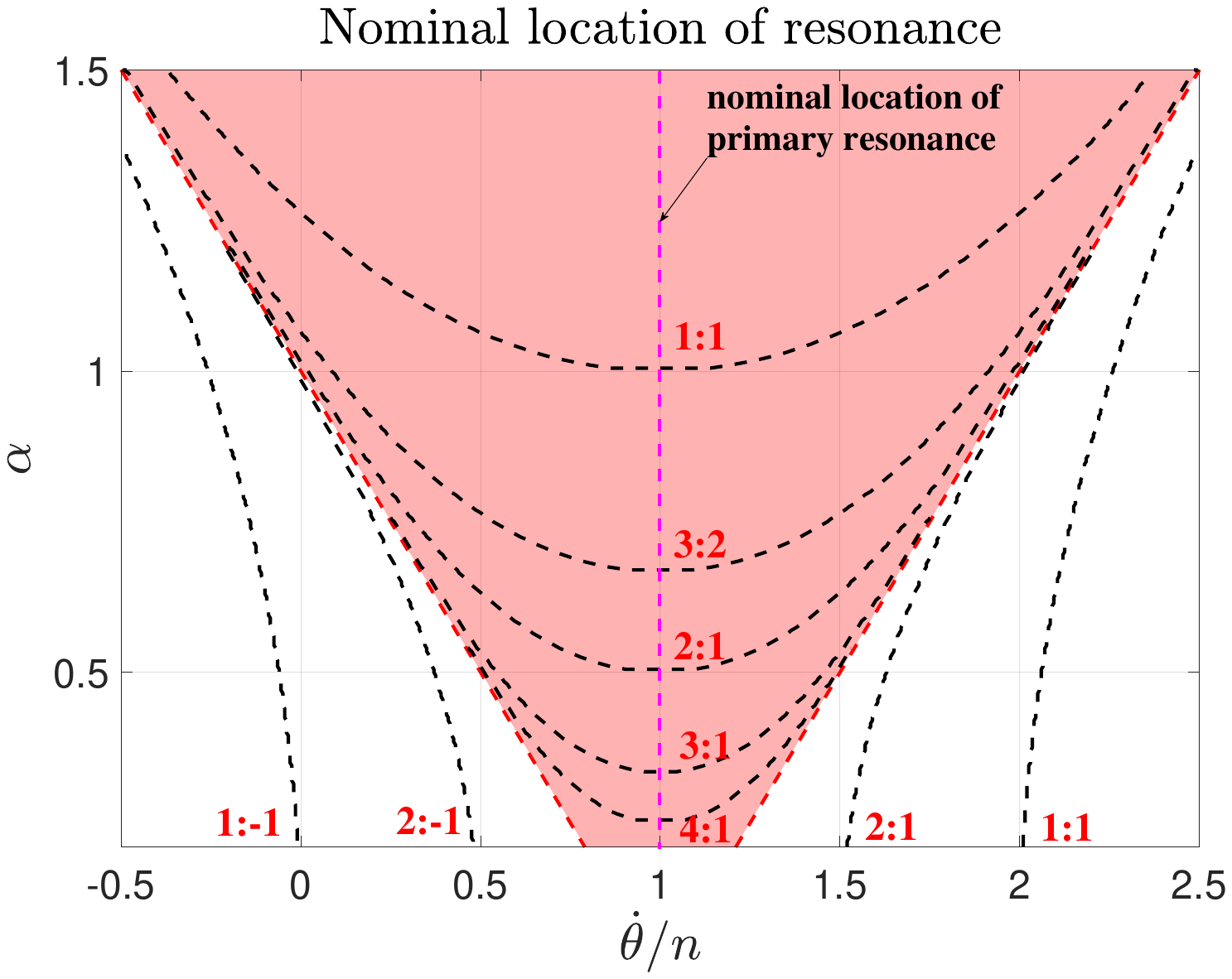}
\includegraphics[width=\columnwidth]{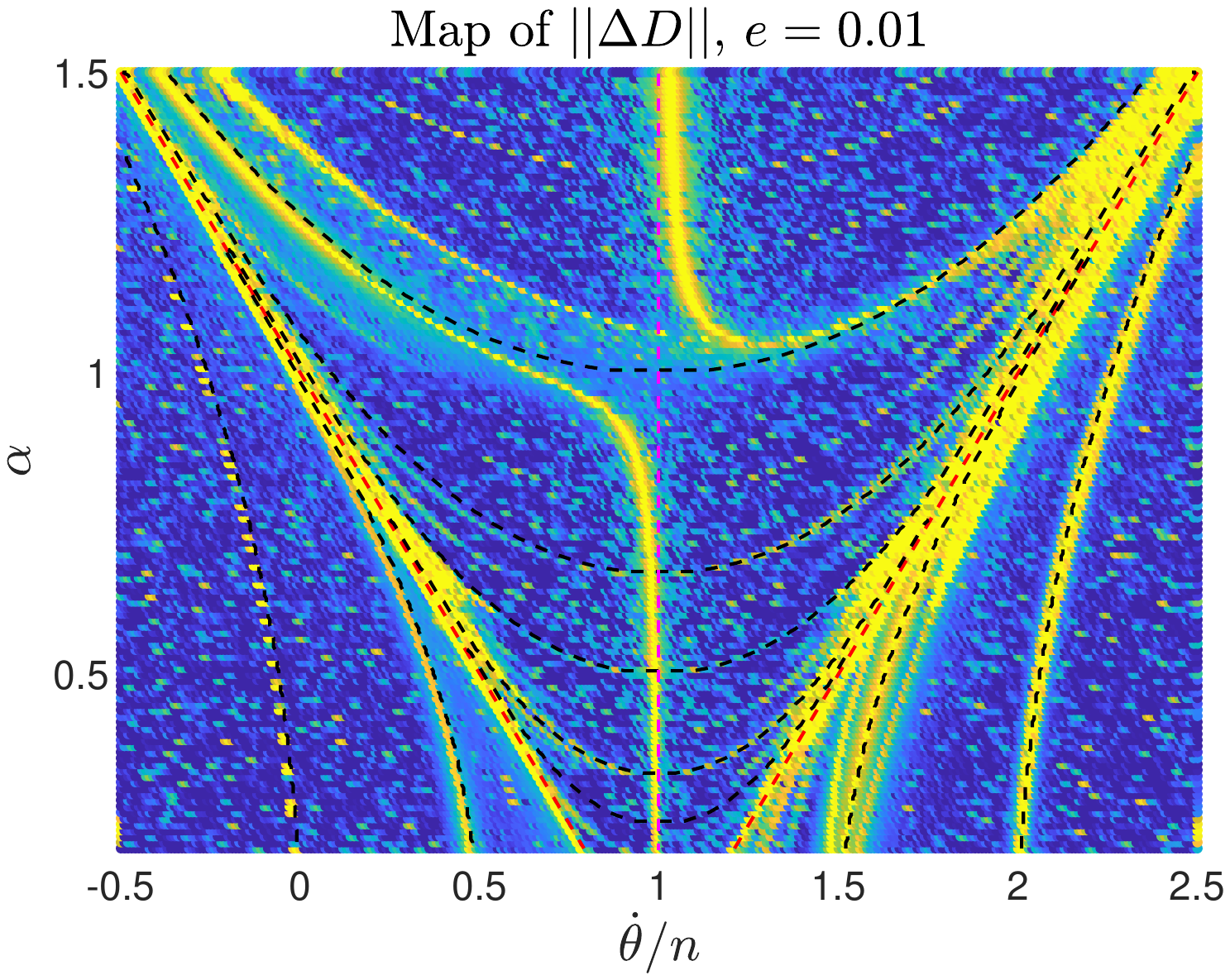}
\caption{Nominal locations of high-order and secondary spin-orbit resonances determined under the Kernel Hamiltonian model shown in the $(\dot \theta, \alpha)$ space (\textit{left panel}) and comparisons to dynamical maps of the second-derivative-based index in the case of $e=0.01$ (\textit{right panel}). The map shown in the right panel is the same as that presented in the left panel of Fig. \ref{Fig2}. In the left panel, the shaded region stands for the parameter space where the synchronous primary resonance happens. Those resonances inside the primary resonance are called secondary resonances and the ones outside the primary resonance are referred to as high-order spin-orbit resonances.}
\label{Fig4}
\end{figure*}

In Fig. \ref{Fig4}, we further present the distribution of high-order and secondary resonance in the $(\dot\theta,\alpha)$ space and compare them to the dynamical map of the normalized second-derivative-based index $||\Delta D||$. The map of $||\Delta D||$ shown in the right panel is the same as the one shown in Fig. \ref{Fig2}. In the left panel, the red shaded region stands for the space where the synchronous primary resonance takes place and the red dashed lines stand for the separatrices of the primary resonance. Secondary resonances happen inside the shaded region and high-order resonances takes place outside the shaded region. It is observed that the nominal locations of spin-orbit resonances are symmetric with respect to the nominal location of the primary resonance at $\dot\theta = 1$. 

From the right panel of Fig. \ref{Fig4}, we can observe that the distributions of high-order and secondary resonances determined under the unperturbed Hamiltonian model are in excellent agreement with the structures arising in the map of $||\Delta D||$.

\subsection{Resonant Hamiltonian model}
\label{Sect4-2}

From the previous subsection, it is known that high-order and secondary resonances may take place in the phase space. The purpose of this section is to formulate a Hamiltonian model for describing the resonant dynamics.

Under the transformation given by equation (\ref{Eq11}), the Hamiltonian can be re-organized as
\begin{equation}\label{Eq15}
{\cal H} = {{\cal H}_0}\left( {\Sigma _1^*,\Sigma _2^*} \right) + {{\cal H}_1}\left( {\sigma _1^*,\sigma _2^*,\Sigma _1^*,\Sigma _2^*} \right),
\end{equation}
where the perturbation part ${\cal H}_1$ is much smaller than the Kernel Hamiltonian ${\cal H}_0$. Therefore, it is reasonable to take ${\cal H}_1$ as a perturbation. 

It should be mentioned that the core concept of henrard's perturbative treatment \citep{henrard1986perturbation, henrard1990semi} lies in transforming the Hamiltonian form given by equation (\ref{Eq9}) to the standard form of Hamiltonian given by equation (\ref{Eq15}). In the standard Hamiltonian, the kernel Hamiltonian is of normal form. The standard Hamiltonian form given by equation (\ref{Eq15}) is ready for studying resonances between $\sigma _1^*$ and $\sigma _2^*$.

In order to study the $k_1$:$k_2$ resonance between $\sigma_1^*$ and $\sigma_2^*$, let us introduce the following transformation,
\begin{equation}\label{Eq16}
\begin{aligned}
{\gamma _1} &= \sigma _1^{\rm{*}} - \frac{{{k_2}}}{{{k_1}}}\sigma _2^{\rm{*}},\quad {\Gamma _1} = \Sigma _1^*,\\
{\gamma _2} &= \sigma _2^{\rm{*}},\quad {\Gamma _2} = \Sigma _2^* + \frac{{{k_2}}}{{{k_1}}}\Sigma _1^*,
\end{aligned}
\end{equation}
which is canonical with the generating function
\begin{equation*}
S = \sigma _1^{\rm{*}}{\Gamma _1} + \sigma _2^{\rm{*}}\left( {{\Gamma _2} - \frac{{{k_2}}}{{{k_1}}}{\Gamma _1}} \right).
\end{equation*}
The resonant argument for the  $k_1$:$k_2$ resonance is ${\gamma _1} = \sigma _1^{\rm{*}} - \frac{{{k_2}}}{{{k_1}}}\sigma _2^{\rm{*}}$.

Under the new set of canonical variables given by equation (\ref{Eq16}), the Hamiltonian can be further expressed as
\begin{equation}\label{Eq17}
{\cal H} = {{\cal H}_0}\left( {{\Gamma _1},{\Gamma _2}} \right) + {{\cal H}_1}\left( {{\gamma _1},{\gamma _2},{\Gamma _1},{\Gamma _2}} \right).
\end{equation}
In particular, when the triaxial satellite is located inside a certain $k_1$:$k_2$ spin-orbit resonance, the corresponding resonant argument $\gamma_1 = \sigma _1^{\rm{*}} - \frac{{{k_2}}}{{{k_1}}}\sigma _2^{\rm{*}}$ becomes a long-periodic angle and the other one $\gamma_2$ is a short-periodic angle. It means that equation (\ref{Eq17}) determines a separable Hamiltonian system \citep{henrard1990semi}. As a consequence, in the long-term evolution, those short-periodic effects can be filtered out by means of averaging theory (corresponding to the lowest-order perturbation theory). To this end, we perform average for the Hamiltonian over $k_1$ times the period of $\gamma_2$, leading to the resonant Hamiltonian,
\begin{equation}\label{Eq18}
\begin{aligned}
{{\cal H}^*} &= \frac{1}{{2{k_1}\pi }}\int\limits_0^{2{k_1}\pi } {{\cal H}\left( {{\gamma _1},{\gamma _2},{\Gamma _1},{\Gamma _2}} \right){\rm d}{\gamma _2}} \\
&= {{\cal H}_0}\left( {{\Gamma _1},{\Gamma _2}} \right) + \frac{1}{{2{k_1}\pi }}\int\limits_0^{2{k_1}\pi } {{{\cal H}_1}\left( {{\gamma _1},{\gamma _2},{\Gamma _1},{\Gamma _2}} \right){\rm d}{\gamma _2}}. 
\end{aligned}
\end{equation}
Under the resonant Hamiltonian model determined by ${\cal H}^*$, the angle $\gamma_2$ becomes a cyclic variable, so that its conjugate momentum
\begin{equation*}
{\Gamma _2} = \Sigma _2^* + \frac{{{k_2}}}{{{k_1}}}\Sigma _1^*
\end{equation*}
is a motion integral. As a result, the resonant Hamiltonian model determined by equation (\ref{Eq18}) is integrable. 

Given the motion integral, the global dynamical structures can be explored by plotting level curves of resonant Hamiltonian ${\cal H}^*$ in the phase space (i.e., phase portraits). In order to compare phase portraits with Poincar\'e sections, we need to project the phase space $(\sigma_1^*,\Sigma_1^*)$ to the space $(\theta,\dot\theta)$ by considering the condition of Poincar\'e sections: ${\rm mod}\left(q_2,2\pi\right)=0$.

\begin{figure*}
\centering
\includegraphics[width=0.9\columnwidth]{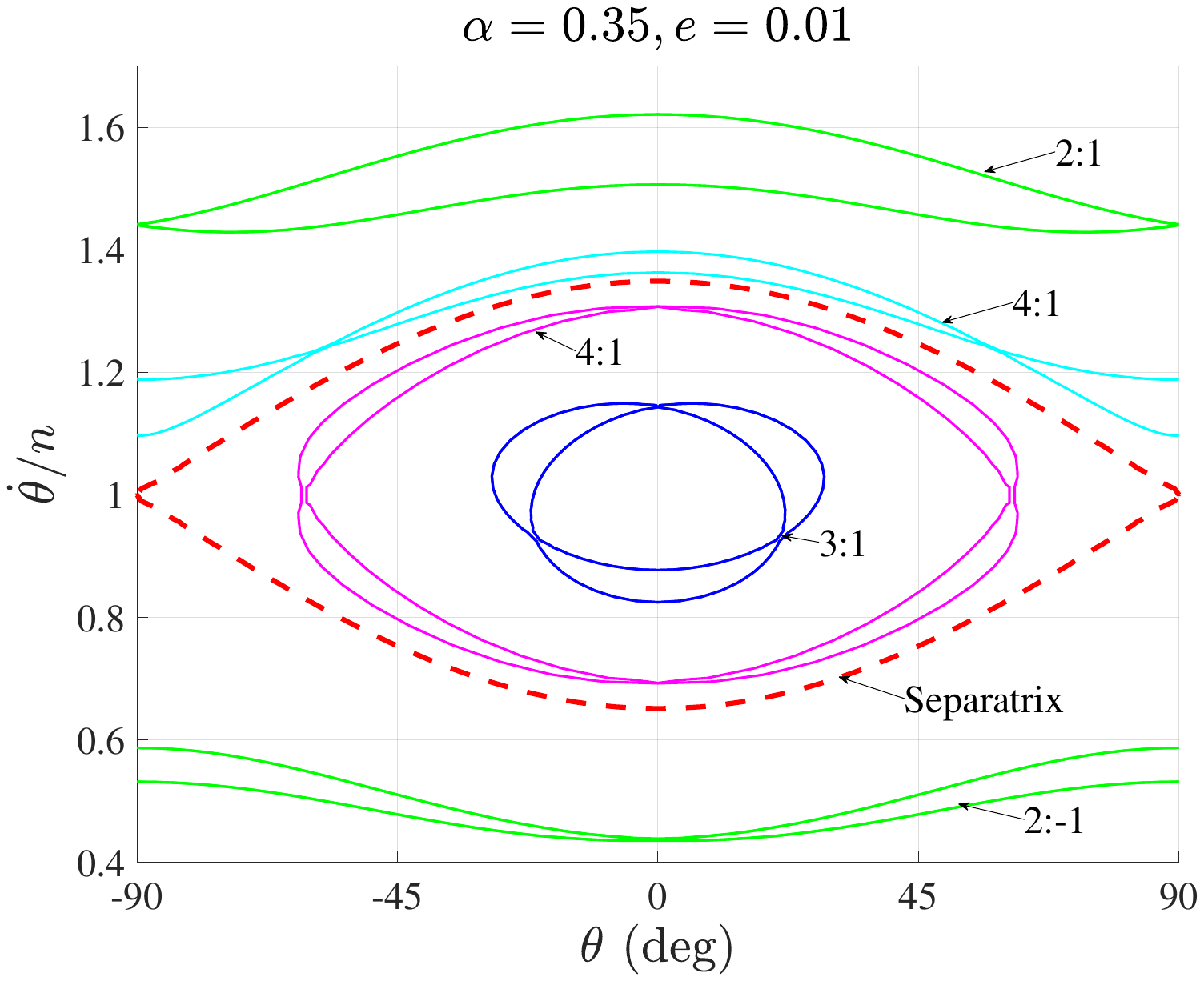}
\includegraphics[width=0.9\columnwidth]{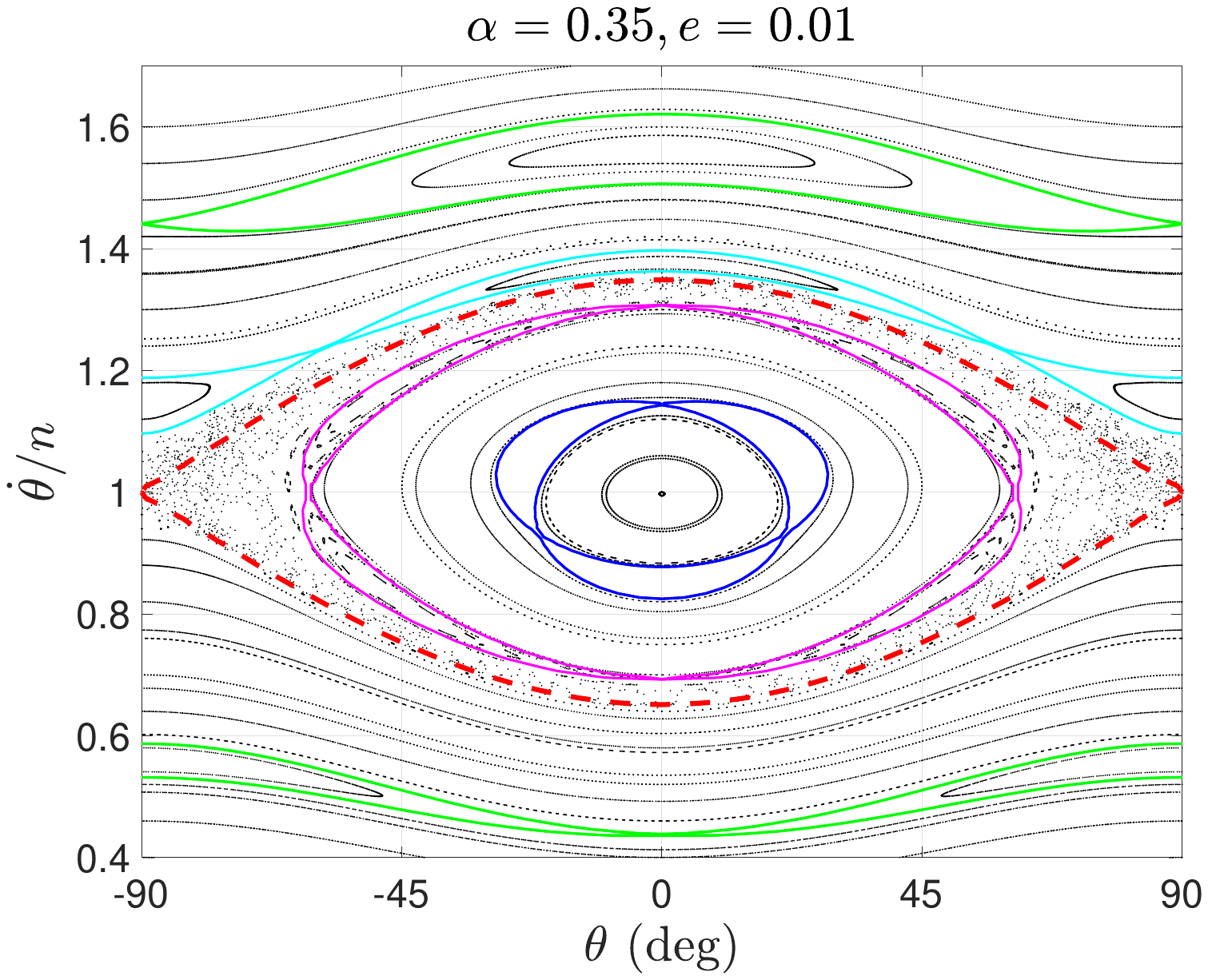}\\
\includegraphics[width=0.9\columnwidth]{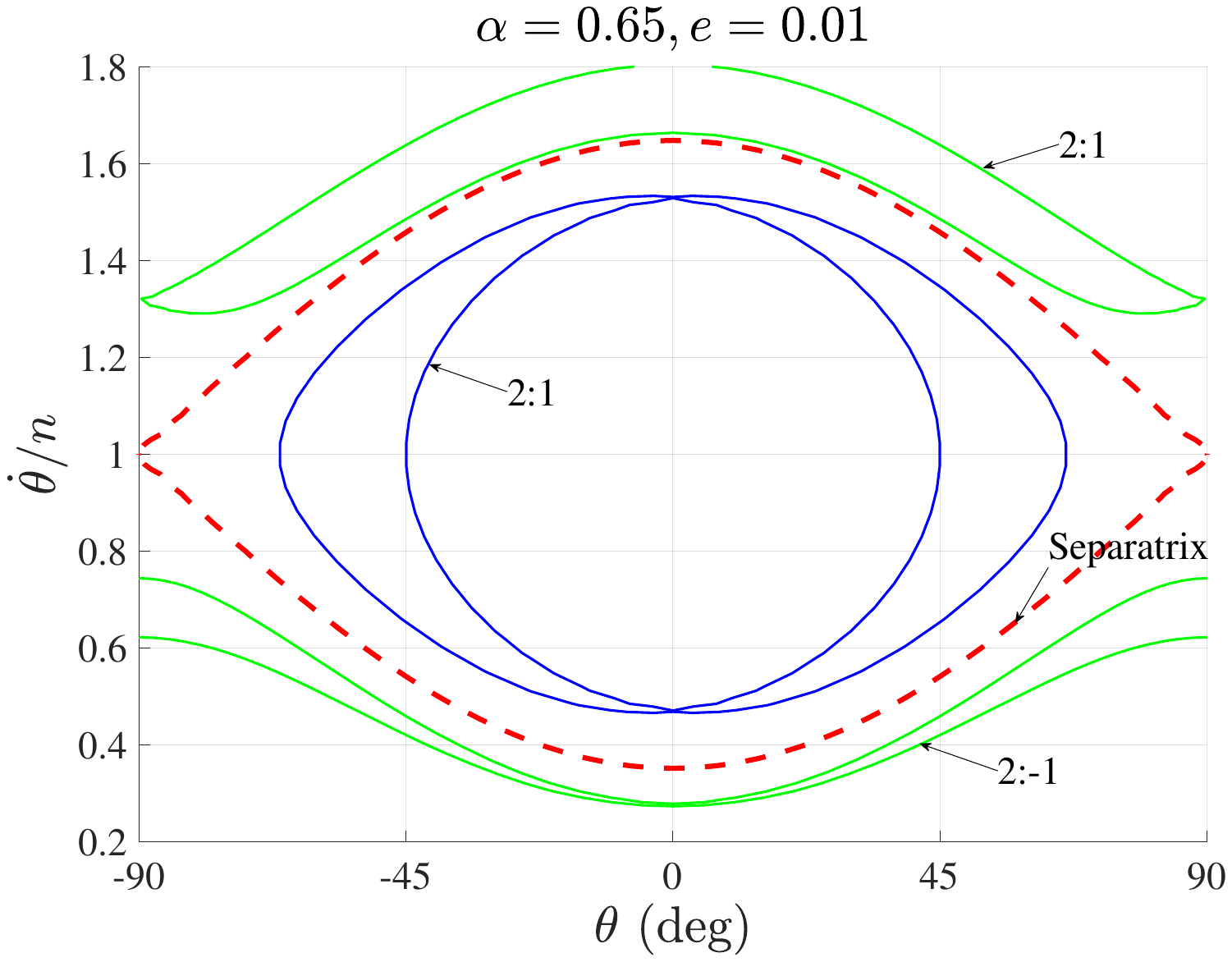}
\includegraphics[width=0.9\columnwidth]{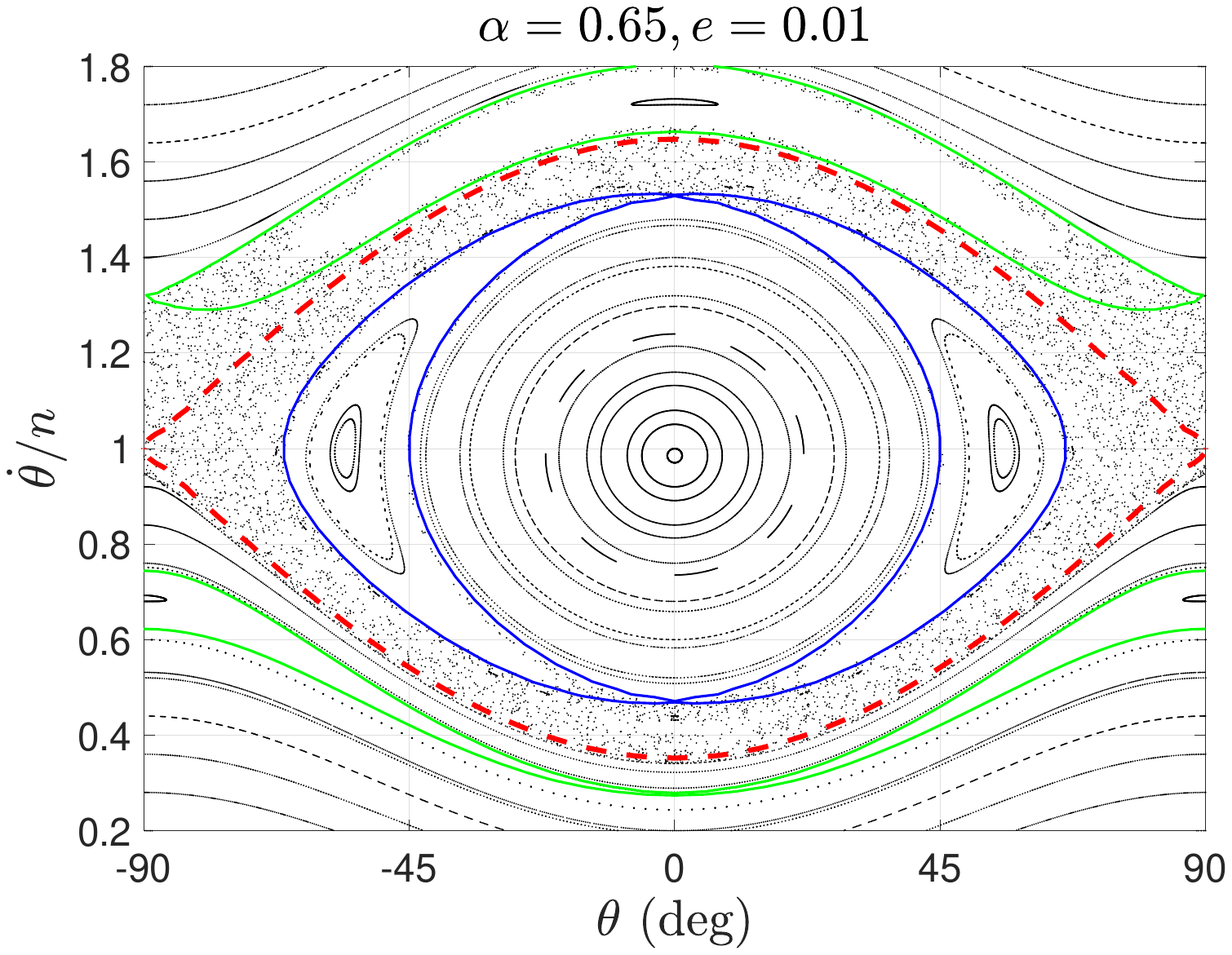}\\
\includegraphics[width=0.9\columnwidth]{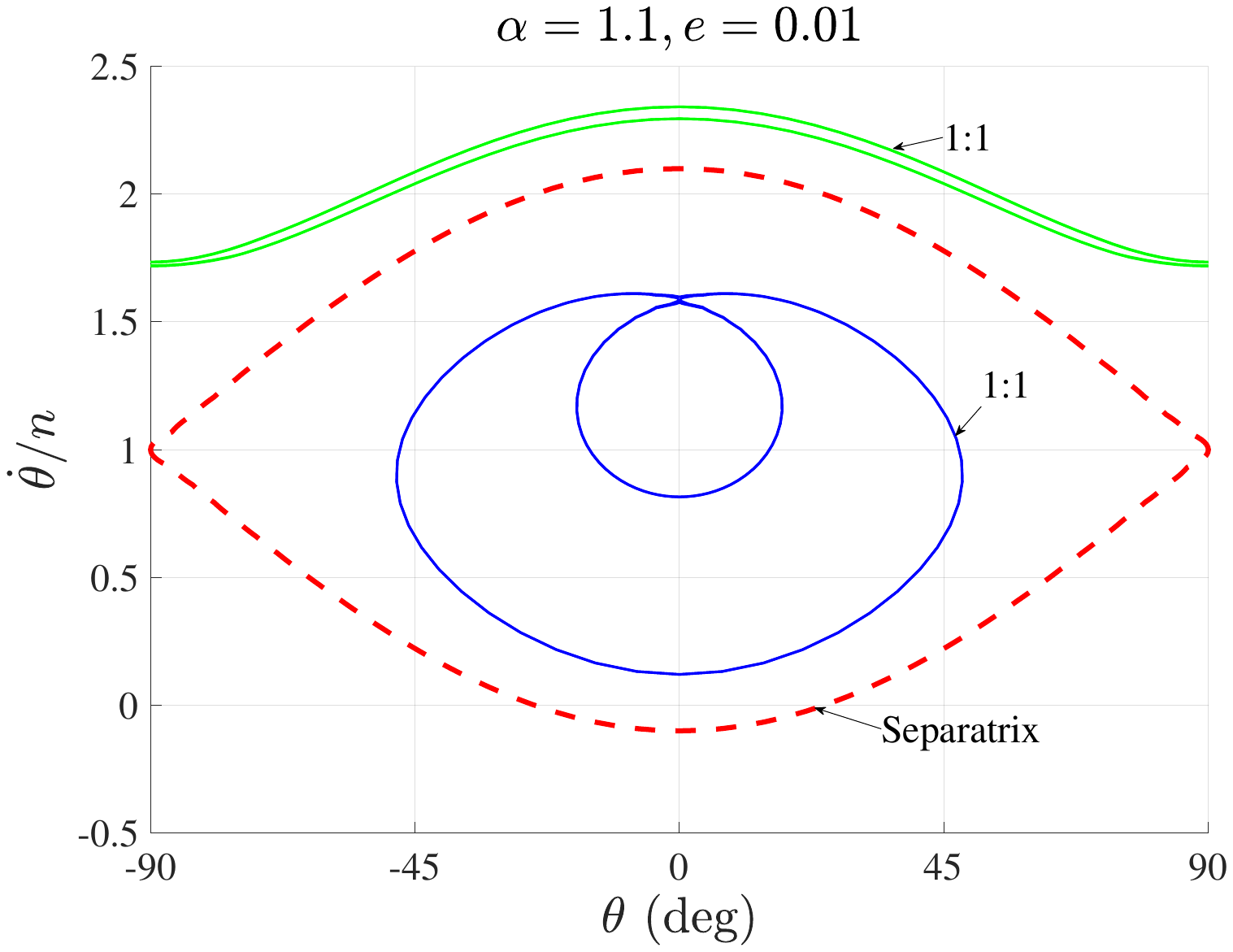}
\includegraphics[width=0.9\columnwidth]{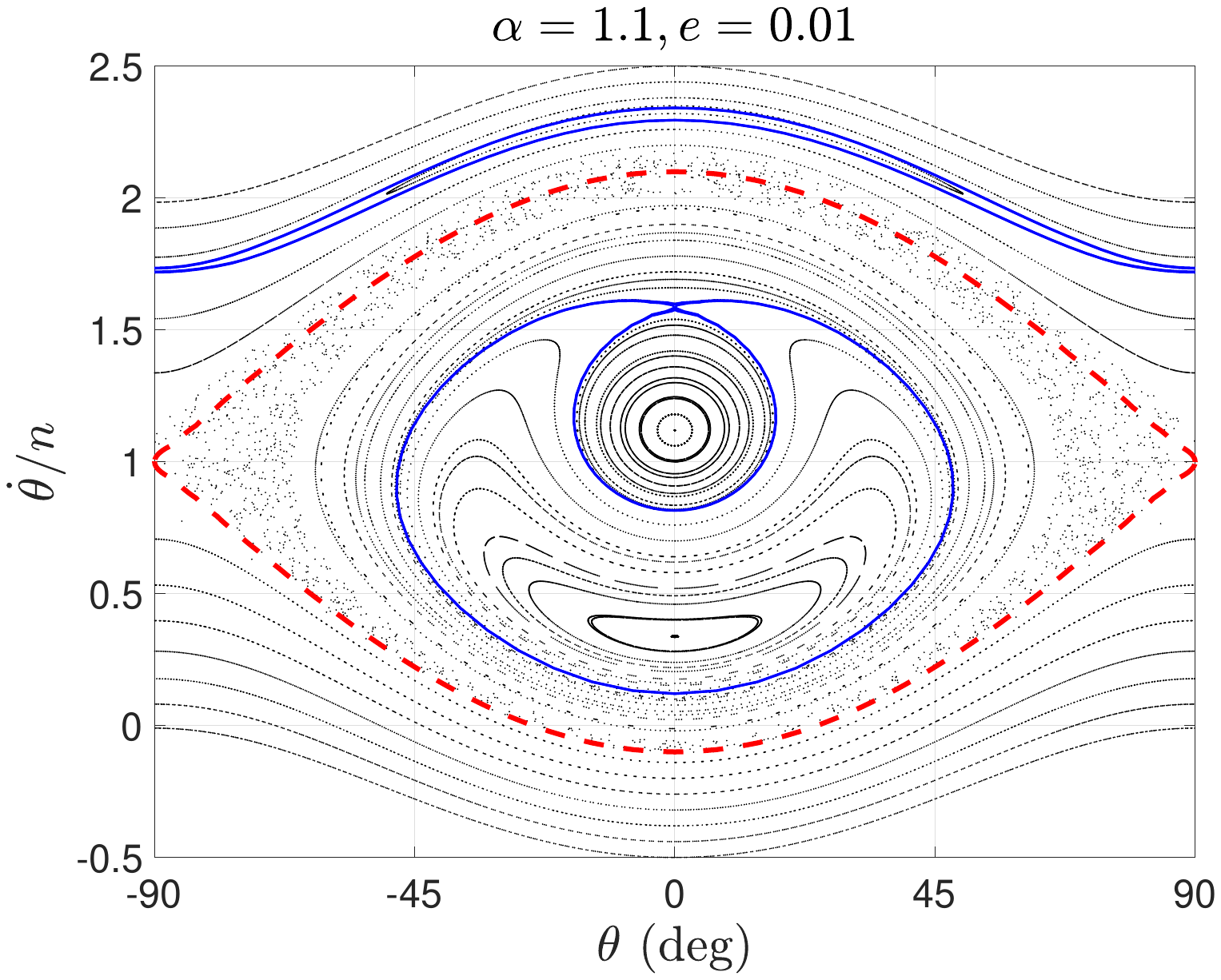}\\
\includegraphics[width=0.9\columnwidth]{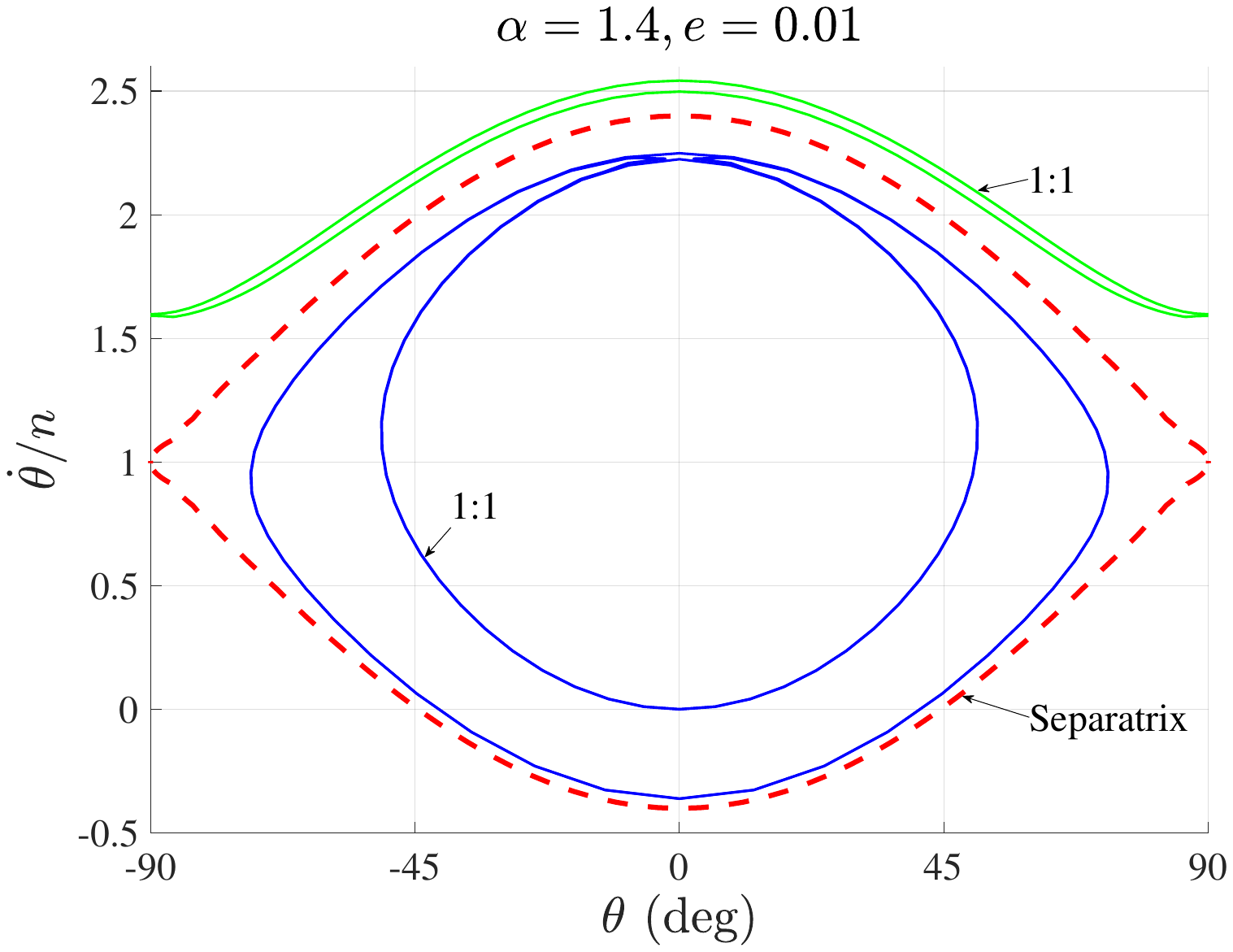}
\includegraphics[width=0.9\columnwidth]{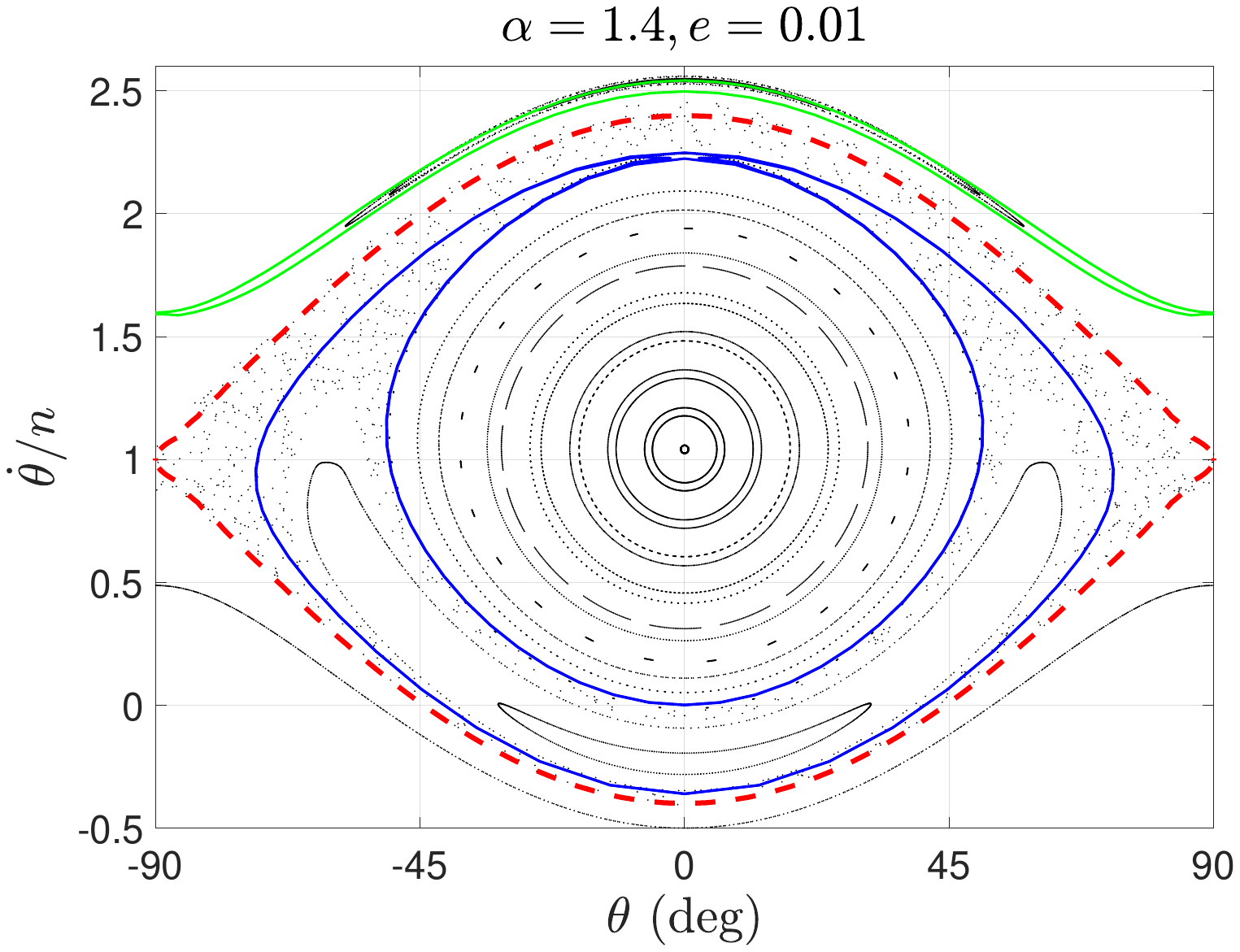}
\caption{Dynamical separatrices of high-order and secondary spin-orbit resonances in the phase space (\textit{left-column panels}) as well as their comparisons to numerical structures arising in Poincar\'e sections (\textit{right-column panels}) for spin-orbit problems with different asphericity parameters. The red dashed lines stand for dynamical separatrices of the synchronous primary resonance, which are expressed by equation (\ref{Eq10}).}
\label{Fig5}
\end{figure*}

\subsection{Results}
\label{Sect4-3}

Analytical and numerical results for high-order and secondary spin-orbit resonances in the phase space $(\theta,\dot\theta)$ are presented in Fig. \ref{Fig5}. Regarding analytical result, we provide the dynamical separatrix, which stands for the isolines of resonant Hamiltonian. Dynamical separatrix divides the phase space into regions of circulation and libration. In essence, dynamical separatrix corresponds to invariant manifolds stemming from unstable periodic orbits in the original dynamical model. In the left-column panels, dynamical separatrices of the synchronous primary resonance, high-order resonances and secondary resonances are shown and, in the right-column panels, they structures are compared to numerical structures arising in Poincar\'e sections. In practical simulations, the spin-orbit problems with asphericity parameters at $\alpha = 0.35$, $\alpha = 0.65$, $\alpha = 1.1$ and $\alpha = 1.4$ are considered and the eccentricity is taken as $e=0.01$. In all panels, the dynamical separatrices of the synchronous primary resonance are marked in red dashed lines.

From Fig. \ref{Fig5}, it is observed that the analytical structures (i.e., dynamical separatrices) of high-order and/or secondary spin-orbit resonances are in good agreement with numerical structures arising in the Poincar\'e sections. In particular, the numerical structures are dominated by the secondary 3:1 and 4:1 spin-orbit resonances and high-order resonances including 2:1, 4:1 and 2:$-1$ in the case of $\alpha = 0.35$, by the secondary 2:1 resonance and high-order resonances including 2:1 and 2:$-1$ in the case of $\alpha = 0.65$ and by the secondary 1:1 resonance and high-order 1:1 resonance in the cases of $\alpha = 1.1$ and $\alpha = 1.4$.

In the spin-orbit problem with $\alpha = 0.35$, the dynamical separatrices of the secondary 4:1 resonance (inside the primary resonance) and high-order 4:1 resonance (outside the primary resonance) are close to the separatrices of the primary resonance, and they provide boundaries for the chaotic layer around the separatrix of the primary resonance. When the asphericity parameter is increased to $\alpha=0.65$, the high-order 4:1 resonance disappears and the secondary 3:1 and 4:1 resonances are replaced by the secondary 2:1 resonance. It is observed that the dynamical separatrices of the high-order 2:1 and secondary 2:1 resonances provide boundaries for chaotic layer around the separatrix of the primary resonance. When the asphericity parameter is further increased up to $\alpha = 1.1$ and $\alpha=1.4$, only high-order 1:1 and secondary 1:1 resonances can be observed in the phase space.

\begin{figure*}
\centering
\includegraphics[width=\columnwidth]{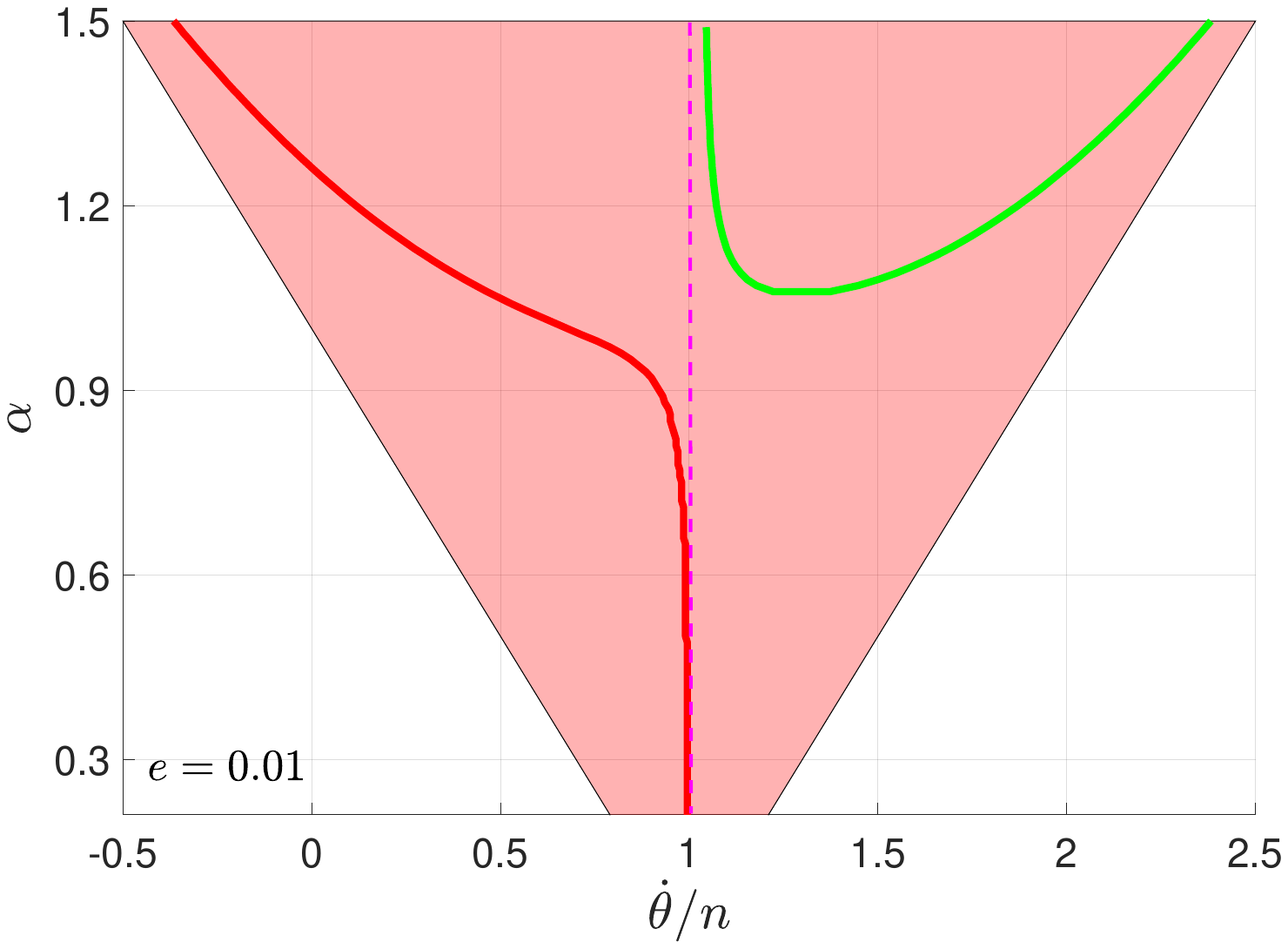}
\includegraphics[width=\columnwidth]{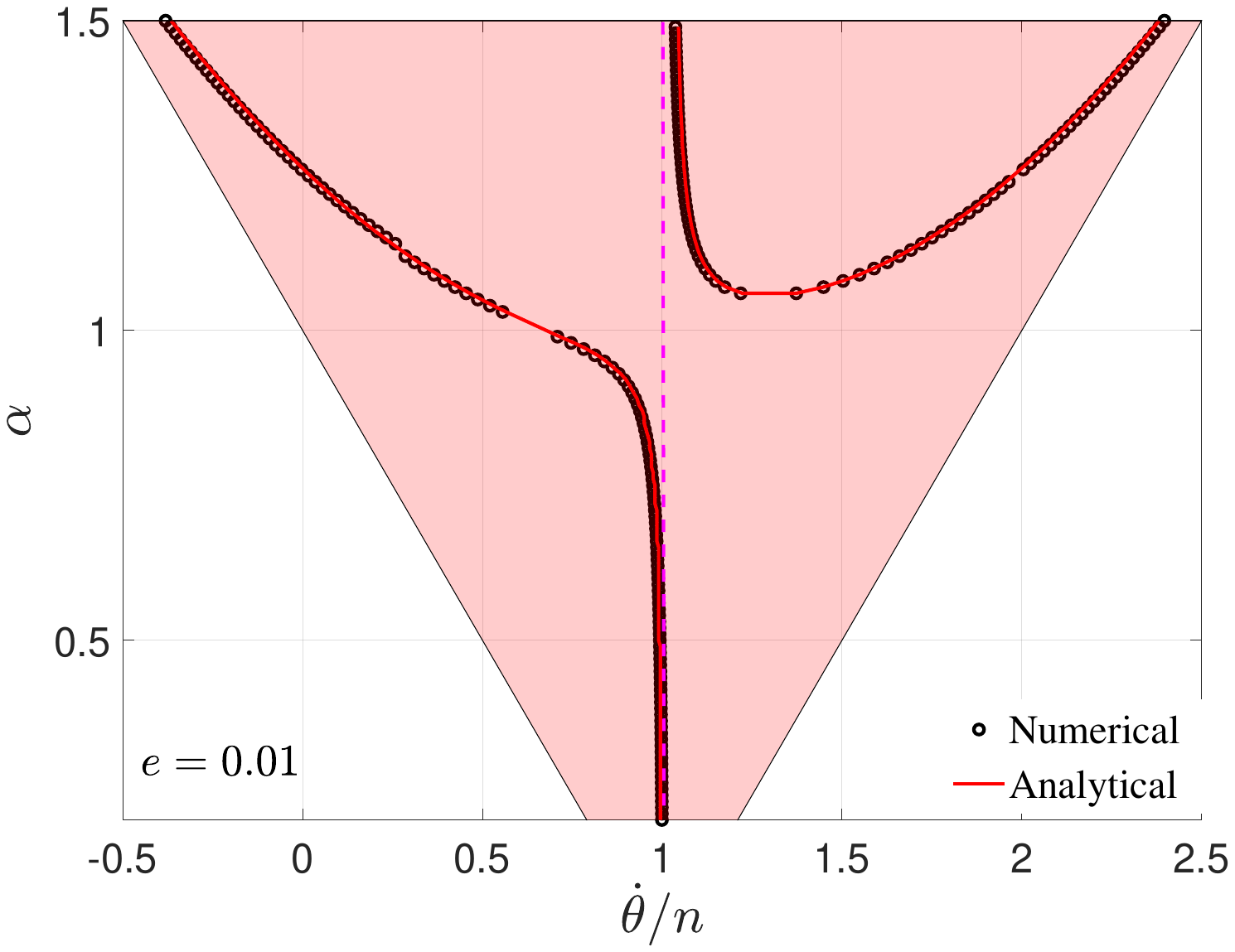}
\caption{Analytical characteristic curves of the synchronous primary resonance and the secondary 1:1 spin-orbit resonance (\textit{left panel}) and comparisons to the associated numerical curves (\textit{right panel}). The vertical dashed line stands for the nominal location of the synchronous primary resonance and the shaded area represents the space where the synchronous primary resonance takes place. The agreement between analytical and numerical curves is good over the entire considered range $\alpha \in [0,1.5]$.}
\label{Fig6}
\end{figure*}

According to the resonant Hamiltonian model, we can analytically determine the location of synchronous primary and/or secondary 1:1 resonances (characteristic curves). On the other hand, the numerical characteristic curves can be determined by computing the associated periodic orbits in a high accuracy. Denote the period by $T$. Periodic orbits should satisfy the following condition:
\begin{equation*}
\left\{ \begin{array}{l}
\theta \left( {t = {t_0}} \right) = \theta \left( {t = {t_0} + T} \right),\\
\dot \theta \left( {t = {t_0}} \right) = \dot \theta \left( {t = {t_0} + T} \right).
\end{array} \right.
\end{equation*}
Without loss of generality, we assume the angle at $t=t_0$ is equal to zero, i.e. ${\theta}_0 = 0$. Thus, there are only two variables $({\dot\theta}_0, T)$ to be determined for a periodic orbit. In practice, we adopt the Newton-Raphson method to solve the unknown variables of periodic orbits.

Fig. \ref{Fig6} shows the analytical characteristic curves in the left panel and comparisons between analytical and numerical curves in the right panel. The shaded region stands for the space where the synchronous primary resonance happens. From Fig. \ref{Fig6}, we can see that (a) the analytical characteristic curves can agree well with the numerical ones in the entire considered range $\alpha \in [0,1.5]$, and (b) the secondary 1:1 resonance appears when the asphericity parameter is higher than $\alpha$$\sim$$1.06$ in the case of $e=0.01$.

Regarding spin-orbit problem, \citet{gkolias2016theory, gkolias2019accurate} took advantage of Lie-series transformation theory to study the low-order secondary resonances bifurcating from the synchronous primary resonance. Analytical characteristic curves are produced from the 11th normal form construction as well as from wisdom formula derived from nonlinear method of Bogoliubov and Mitropolsky \citep{wisdom1984chaotic}, and analytical results are compared to numerical computations. It is concluded that the normal form method has a significant improvement compared to the nonlinear method of Bogoliubov and Mitropolsky. In particular, the normal form solution can agree with numerical result in the range of $\alpha\in [0, 1.2]$, but starts to diverge when $\alpha$ is greater than 1.2 (please refer to Fig. 2 in their work for more details). However, our analytical estimate can match the numerical solution in the entire considered range of $\alpha$ up to 1.5 (see the right panel of Fig. \ref{Fig6}).

\begin{figure*}
\centering
\includegraphics[width=\columnwidth]{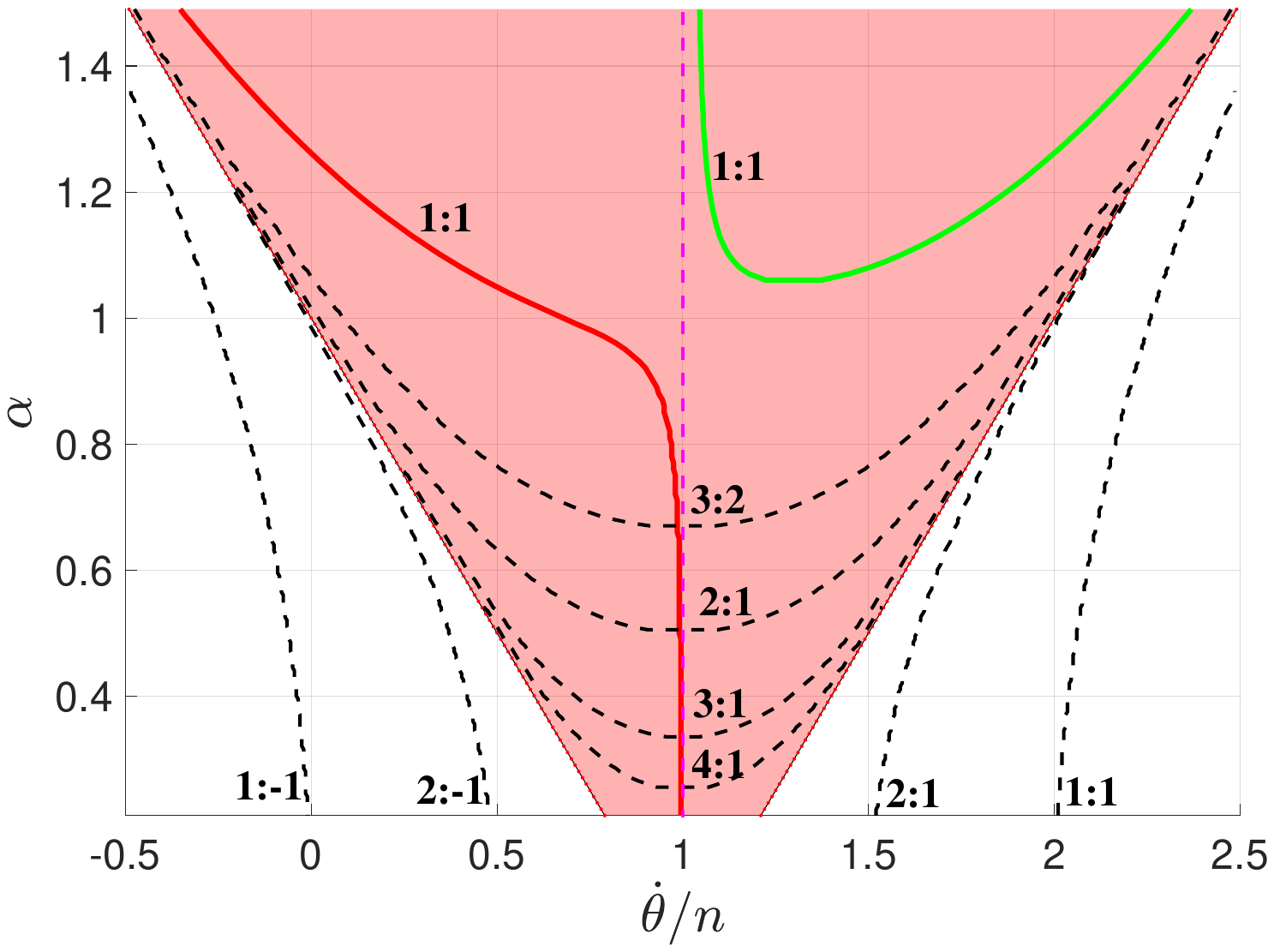}
\includegraphics[width=\columnwidth]{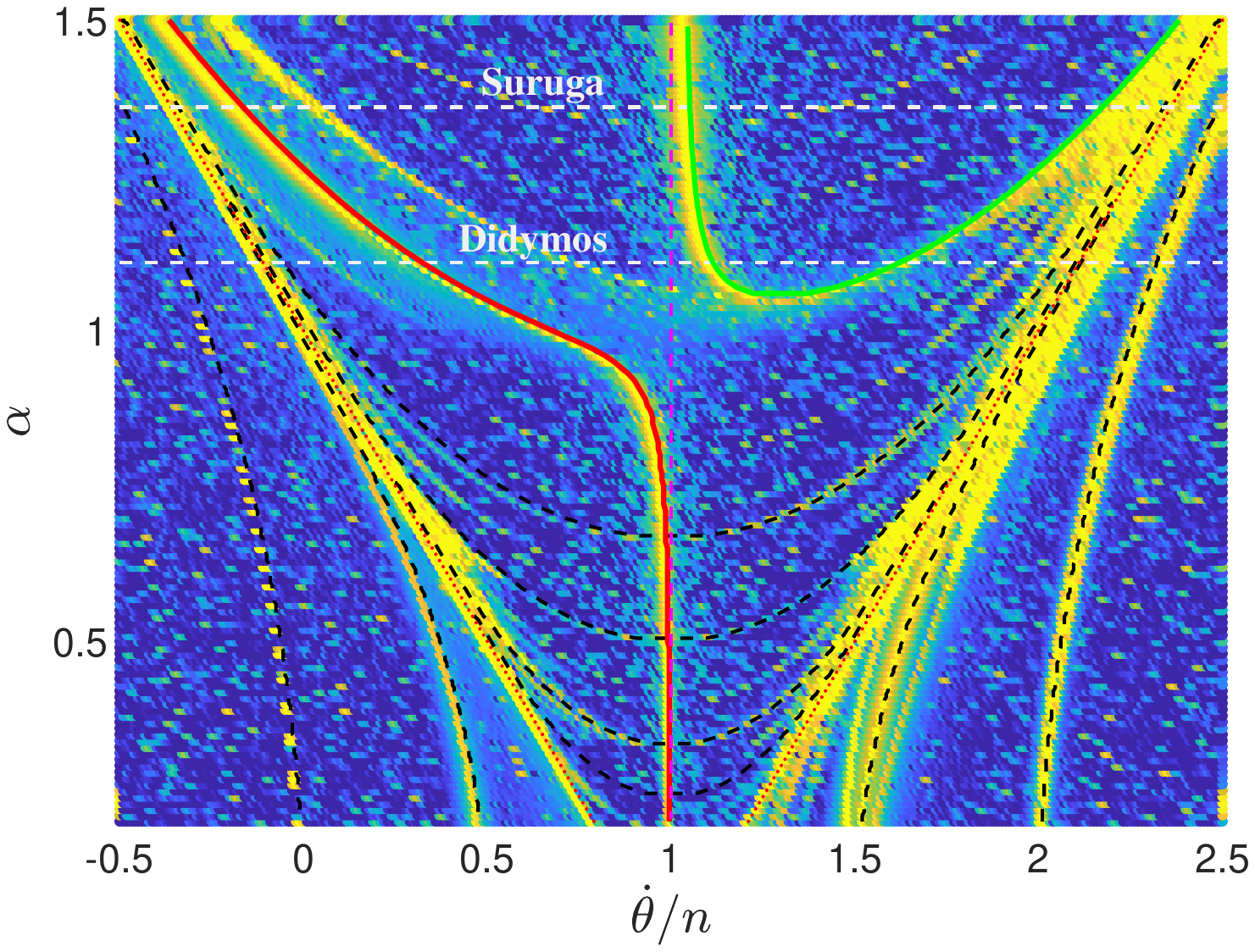}
\caption{Analytical characteristic curves of the synchronous primary resonance, high-order resonances (outside the primary resonance) and secondary resonances (inside the primary resonance) in the $(\dot\theta,\alpha)$ space (\textit{left panel}) and comparisons with dynamical maps of the normalized second-derivative-based index $||\Delta D||$ (\textit{right panel}). In the right panel, the locations of two typical binary-asteroid systems are marked. The asphericity parameter is $\alpha = 1.1$ for the binary-asteroid system (65803) Didymos and it is 1.36 for (4383) Suruga. The vertical dashed line stands for the nominal location of the synchronous primary resonance and the shaded region in the left panel represents the space where the synchronous primary resonance happens.}
\label{Fig7}
\end{figure*}

Fig. \ref{Fig7} shows all the analytical characteristic curves of the synchronous primary resonance, secondary resonances and high-order resonances in the left panel and makes a comparison between analytical curves to the numerical map of $||\Delta D||$ in the right panel. We can see that the structures arising in the dynamical map have a good correspondence to the analytical characteristic curves. Thus, these numerical structures can be well understood thanks to the associated spin-orbit resonance dynamics.

In particular, Fig. \ref{Fig7} provides a global view for the rotational dynamics of the secondary in the $(\dot\theta, \alpha)$ space. It is possible for us to predict possible dynamical outcomes for binary asteroid systems according to the associated asphericity parameters. For example, the following predictions can be made: (a) secondary 1:1 resonance may appear when $\alpha > 1.06$, (b) high-order 2:1 and 2:$-1$ resonances as well as secondary 2:1, 3:1, and 4:1 resonances may disappear in the high-$\alpha$ space because they are inside the chaotic sea around the separatrix of the synchronous primary resonance, and (c) high-order 1:1 and 1:$-1$ resonances can appear in the entire considered space but the strength of high-order 1:$-1$ resonance is relatively weak.

\begin{table*}
\centering
\caption{Physical parameters for the example binary asteroid systems adopted in this work (please refer to http://www.asu.cas.cz/asteroid/binastdata.htm for detailed data of binary asteroid systems). The ratio of spin angular momentum to the orbital angular momentum for the secondary $L_{\rm spin}/L_{\rm orb}$ is estimated from the method given in \citet{jafari2023surfing}.}
\begin{tabular*}{\textwidth}{@{\extracolsep{\fill}}lllllllcccc@{\extracolsep{\fill}}}
\hline
{binary system}&{$D_s/D_p$}&{$m_s/m_p$}&{$P_p$ (h)}&{$P_{\rm orb}$ (h)}&{$P_s$ (h)}&{$a_s/b_s$}&{$a_{\rm orb}/D_s$}&{$e_{\rm orb}$}&{$L_{\rm spin}/L_{\rm orb}$}&{$\alpha$}\\
\hline
(65803) Didymos&0.227&0.012&2.2593&11.91&11.91&1.55&6.706&$<0.03$&$2.2\times10^{-3}$&1.1\\
(4383) Suruga&0.188&0.007&3.4068&16.34&16.34&2.04&10.0&$-$&$9.9\times10^{-4}$&1.36\\
\hline
\end{tabular*}
\label{Tab1}
\end{table*}

\section{Applications to binary asteroid systems}
\label{Sect5}

The analytical approach is applied to two representative binary asteroid systems: (65803) Didymos and (4383) Suruga \citep{pravec2016binary, warner2013something}. (65803) Didymos is the target of the DART mission \citep{agrusa2022dynamical}. It should be mentioned that the analytical method presented in this work is also applicable for other binary asteroid systems. The location of these two systems are marked in the right panel of Fig. \ref{Fig7}, which shows  that their asphericity parameters are greater than the critical value $\alpha_c = 1.06$ at which the secondary 1:1 resonance bifurcates from the synchronous primary resonance. The physical parameters of the adopted binary asteroid systems are shown in Table \ref{Tab1}, including the ratio of the secondary diameter to the primary diameter $D_s/D_p$, ratio of the secondary mass to the primary mass $m_s/m_p$, spin period of the primary $P_p$, mutual orbital period of system $P_{\rm orb}$, spin period of the secondary $P_s$, the equatorial elongation of the secondary $a_s/b_s$, ratio of the orbital semimajor axis to the secondary's diameter $a_{\rm orb}/D_s$, mutual orbital eccentricity $e_{\rm orb}$, ratio of spin angular momentum to the orbital angular momentum $L_{\rm spin}/L_{\rm orb}$ and asphericity parameter $\alpha$. 

Regarding binary asteroid systems, \citet{jafari2023surfing} offered a criterion for the definition of dynamical closeness of asteroid pairs: a binary asteroid system is a close pair if the ratio of rotational angular momentum to the orbital angular momentum is greater than 10 percent. According to their estimate, we can get the ratios of spin angular momentum to the orbital angular momentum are $L_{\rm spin}/L_{\rm orb} = 2.2\times10^{-3}$ for (65803) Didymos and $L_{\rm spin}/L_{\rm orb} = 9.9\times10^{-4}$ for (4383) Suruga, which are much smaller than 10 percent. Thus, both binary asteroid systems are far away from close binaries, which means that the couple between orbital and rotational evolution is very weak. In the dynamical evolution, the influence of rotational evolution upon the orbital dynamics can be ignored. As a result, the spin-orbit problem discussed in this work is a good approximation for such a kind of weakly-coupled binary systems. 

According to Table \ref{Tab1}, the asphericity parameter is $\alpha = 1.1$ for binary asteroid system (65803) Didymos and it is equal to $\alpha = 1.36$ for binary asteroid system (4383) Suruga. Considering that the orbital eccentricities are small and highly uncertain, we take two small eccentricities at $e=0.01$ and $e=0.02$ in practical simulations. 

\begin{figure*}
\centering
\includegraphics[width=\columnwidth]{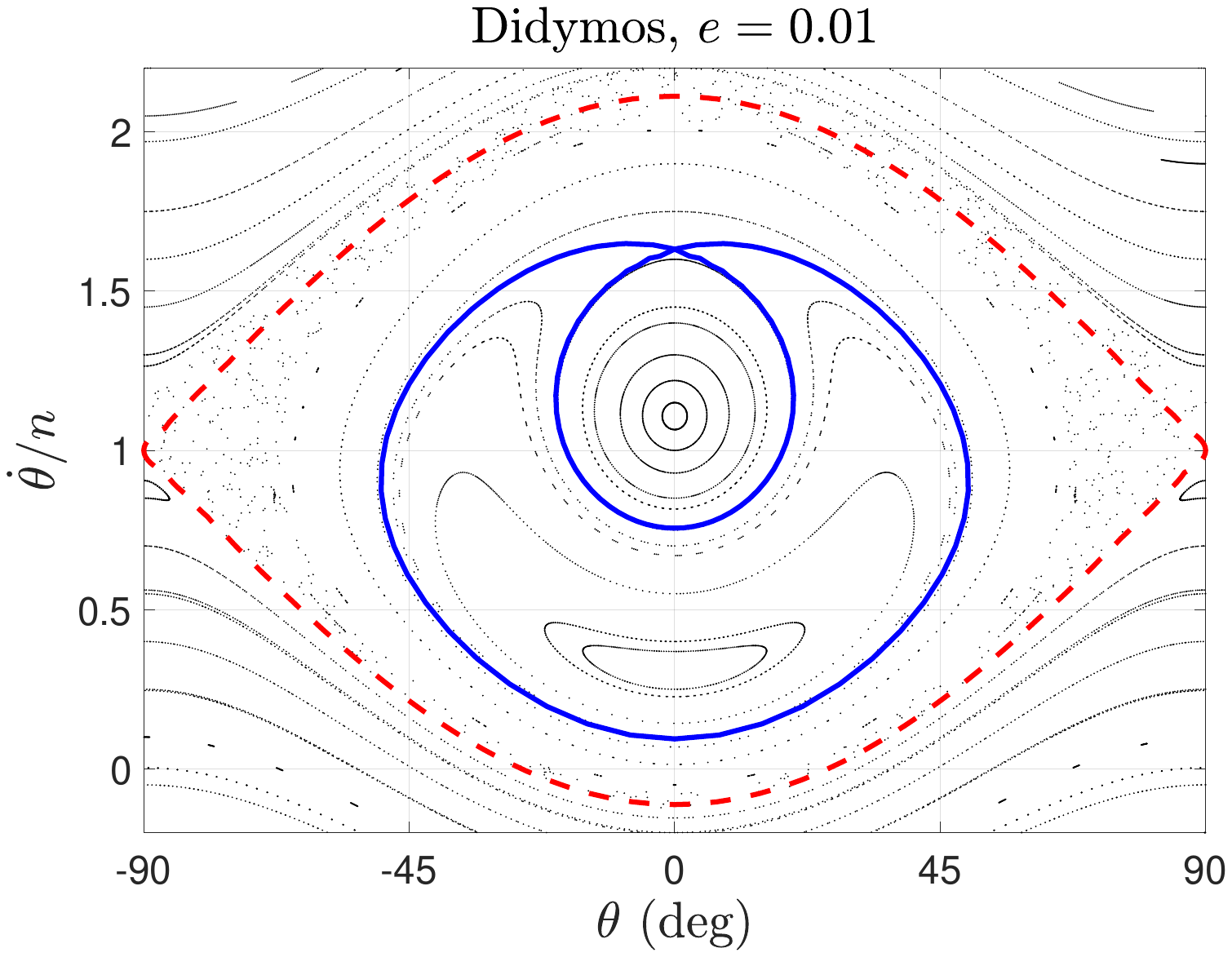}
\includegraphics[width=\columnwidth]{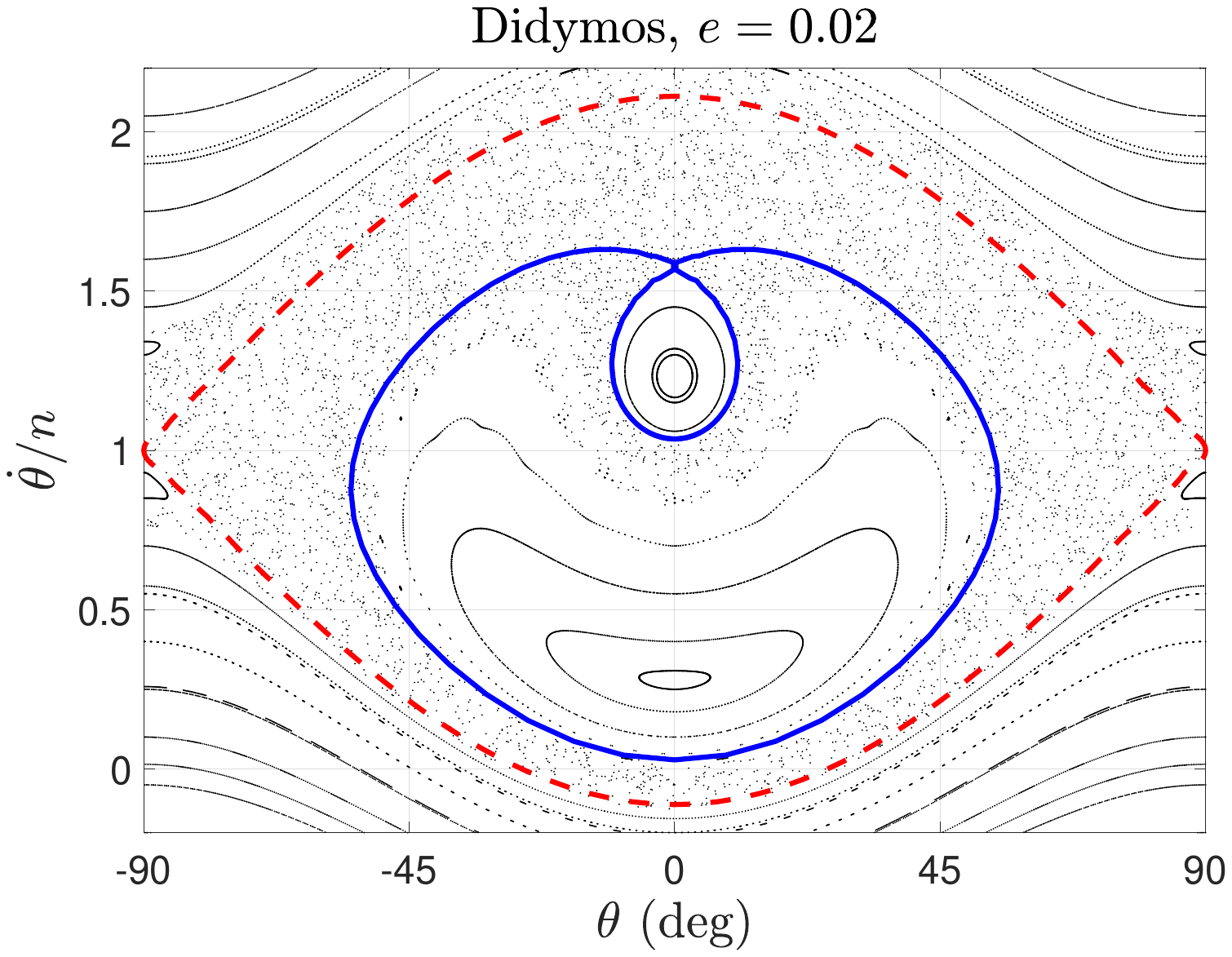}\\
\includegraphics[width=\columnwidth]{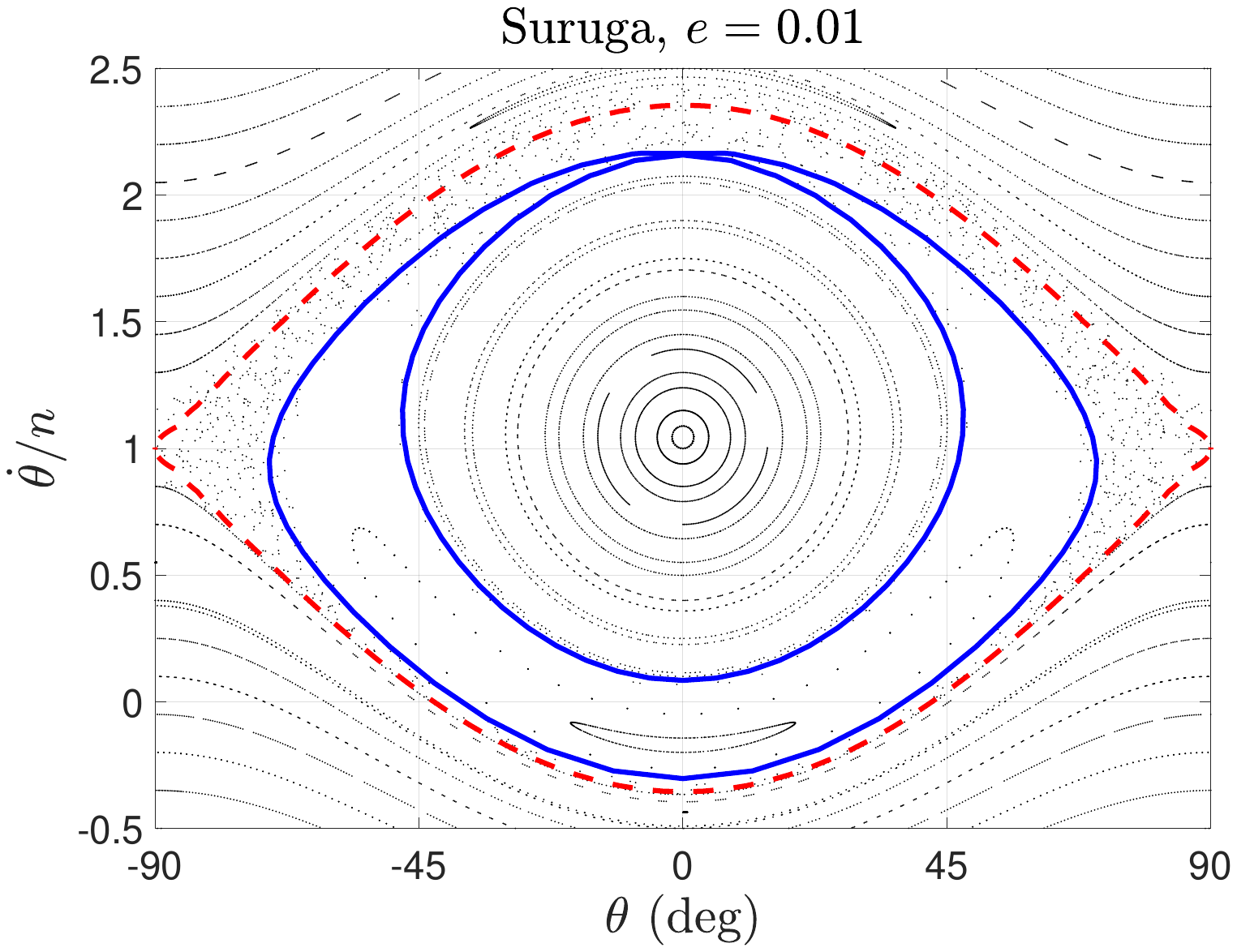}
\includegraphics[width=\columnwidth]{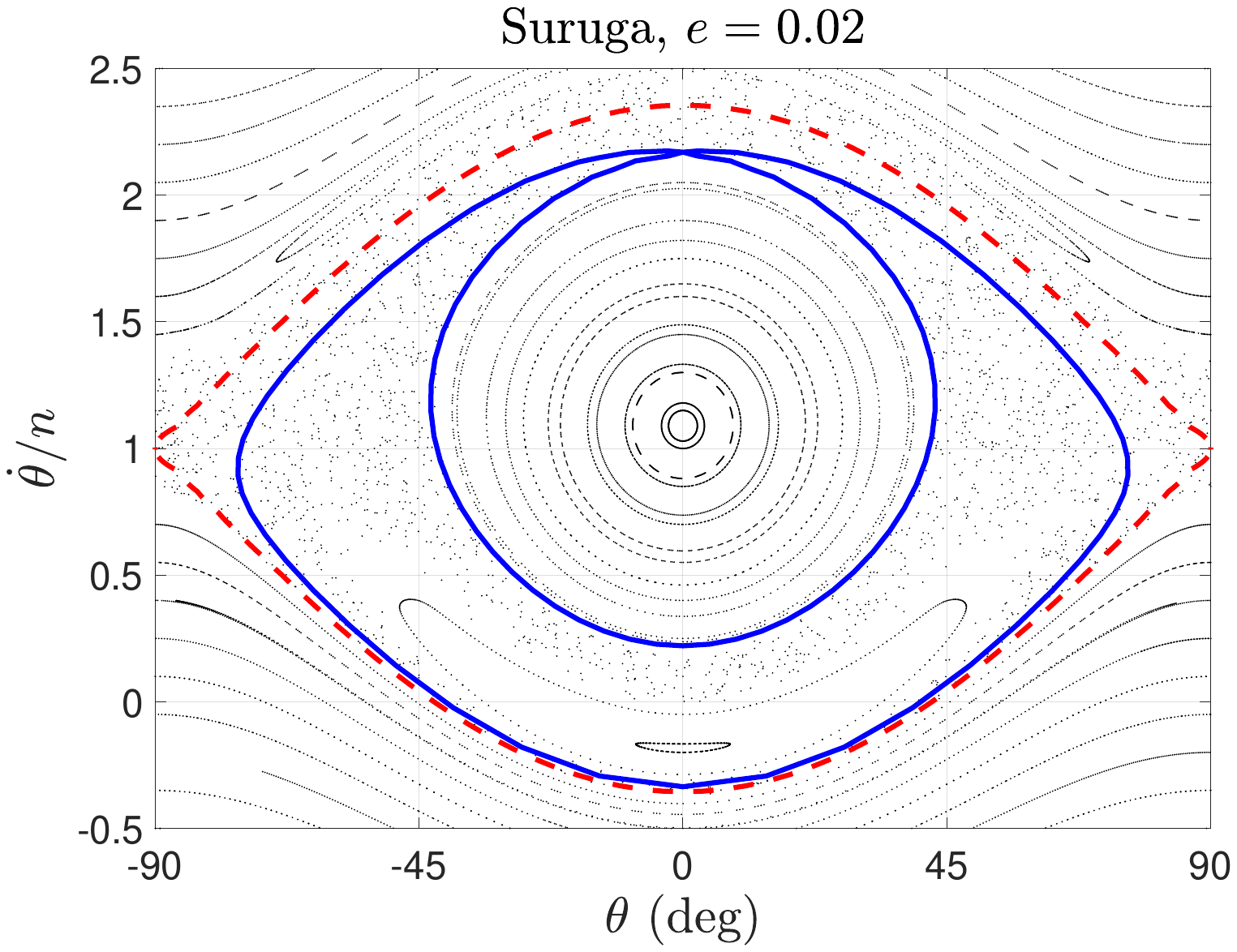}
\caption{Analytical structures (dynamical separatrices) of the synchronous primary resonance and the secondary 1:1 spin-orbit resonance together with the associated numerical structures arising in Poincar\'e sections at different eccentricities for binary asteroid systems (65803) Didymos (\textit{top panels}) and (4383) Suruga (\textit{bottom panels}). The blue lines stand for the dynamical separatrices of the 1:1 secondary resonance and red dashed lines stand for the dynamical separatrices of the synchronous primary resonance. The asphericity parameter is $\alpha = 1.1$ for binary asteroid system (65803) Didymos and it is $\alpha = 1.36$ for binary asteroid system (4383) Suruga.}
\label{Fig8}
\end{figure*}

Analytical and numerical results are presented in Fig. \ref{Fig8}, where the top panels are for (65803) Didymos and the bottom panels are for (4383) Suruga. The dynamical separatrix of the synchronous primary resonance is shown in red dashed line and the separatrix of the secondary 1:1 resonance is given in blue line. From Fig. \ref{Fig8}, we can observe that (a) there is an excellent agreement between the analytical and numerical structures, and (b) chaotic layer becomes wilder with a higher eccentricity. 

According to the current physical parameters, it is highly possible that the binary asteroid systems (65803) Didymos and (4383) Suruga are inside the secondary 1:1 spin-orbit resonance. However, there are a large uncertainty for their elongation $a_s/b_s$, leading to a large uncertainty for determining asphericity parameter $\alpha$. For example, it will be outside the secondary 1:1 resonance if the uncertainty of asphericity parameter makes $\alpha$ be smaller than $\sim$1.06. Thus, more observations are required to further determine the asphericity parameter $\alpha$ accurately in order to predict their real resonant states. However, Fig. \ref{Fig7} can help to predict other possible resonant states when $\alpha$ is varied.

\section{Conclusions}
\label{Sect6}

In this work, high-order and secondary resonances in the spin-orbit problem are investigated. Dynamical maps are numerically produced in order to provide a global picture for the dynamics. In particular, two types of dynamical indicators are introduced. The first one is the fast Lyapunov indicator (FLI) and the second one is the normalized second-derivative-based index $||\Delta D||$. The maps of FLI can distinguish the regular and chaotic regions and the maps of $||\Delta D||$ has an ability to detect minute structures in the phase space. It is concluded that the main V-shape structures are sculpted by the synchronous primary resonance, those minute structures inside the V-shape region are governed by secondary resonances and the ones outside the V-shape region are dominated by high-order spin-orbit resonances. 

The main purpose of this work is to understand dynamical causes of numerical structures. To this end, the theory of perturbative treatments is adopted. In particular, the Hamiltonian is divided into the kernel Hamiltonian and the perturbation part. The terms related to the synchronous primary resonance are included in the kernel Hamiltonian, so that it is possible to consider the influence of the synchronous primary resonance upon high-order and/or secondary resonances. Under the kernel Hamiltonian model, a series of canonical transformations are introduced and nominal locations of high-order and secondary resonances are determined. By taking advantage of Henrard's perturbative treatments \citep{henrard1990semi}, resonant Hamiltonian model is formulated. Thus, it is ready to study high-order and secondary resonance dynamics analytically. Results show that there is an excellent correspondence between analytical structures arising in phase portraits and numerical structures arising both in the dynamical maps and in the Poincar\'e sections. Applications to two binary asteroid systems (including Didymos and Suruga) show that these two binary systems are possible to be inside the secondary 1:1 spin-orbit resonance.

The main results are presented in Fig. \ref{Fig7} where the distributions of synchronous primary resonance, high-order resonance and secondary resonances are presented and compared with dynamical map of $||\Delta D||$. It provides a global picture for us to predict possible resonant states when the asphericity parameter is changed.

\begin{acknowledgments}
This work is supported by the National Natural Science Foundation of China (Nos. 12073011, and 12233003) and the National Key R\&D Program of China (No. 2019YFA0706601).
\end{acknowledgments}



\bibliography{mybib}{}
\bibliographystyle{aasjournal}



\end{document}